\documentclass[letterpaper,twocolumn,10pt]{article}
\PassOptionsToPackage{hyphens}{url}
\usepackage{usenix-2020-09}

\usepackage{tikz}
\usepackage{amsmath}

\usepackage{filecontents}
\usepackage{xcolor}
\usepackage{booktabs}
\usepackage[labelfont=bf]{caption}
\usepackage{graphicx}
\usepackage{multirow}
\usepackage{listings}
\usepackage[compact]{titlesec}
\usepackage{float}
\usepackage{subfig}

\usepackage[done=false, later=false, notes=false]{dtrt}

\newcommand{\roei}[1]{\dtcolornote[Roei]{purple}{#1}}

\definecolor{darkgreen}{rgb}{0,0.6,0}
\newcommand{\eran}[1]{\dtcolornote[Eran]{darkgreen}{#1}}
{}

\newcommand{\paragraphbe}[1]{\vspace{0.75ex}\noindent{\bf \em #1} }

\newcommand{\size}[1]{\left|#1\right|}

\newcommand{\atype}{\mathfrak{b}}
\newcommand{\ofatype}[1]{#1^\atype}

\newcommand{\target}{t}
\newcommand{\features}{\mathcal{F}_\target}

\newcommand{\lline}{l}

\newcommand{\untargeted}[1]{#1_{U}}
\newcommand{\targeted}[1]{#1_{\target}}
\newif\ifsimple
\simpletrue

\ifsimple{}
\newcommand{\pset}{\mathcal{P}}
\newcommand{\psetu}{\pset}
\newcommand{\psett}{\pset}
\newcommand{\goodfile}{\mathcal{G}}
\newcommand{\badfile}{\mathcal{B}}
\newcommand{\badfilet}{\badfile}
\newcommand{\uglyfile}{\mathcal{U}}
\newcommand{\tlines}{\mathcal{T}^{\atype}}
\newcommand{\ulines}{\mathcal{T}^{\mathfrak{u}}}
\else{}
\newcommand{\pset}{\ofatype{\mathcal{P}}}
\newcommand{\psetu}{\untargeted{\pset}}
\newcommand{\psett}{\targeted{\pset}}
\newcommand{\goodfile}{\ofatype{\mathcal{G}}}
\newcommand{\badfile}{\ofatype{\mathcal{B}}}
\newcommand{\badfilet}{\targeted{\badfile}}
\newcommand{\uglyfile}{\ofatype{\mathcal{U}}}
\newcommand{\tlines}{\ofatype{\mathcal{T}}}
\fi{}

\usepackage[utf8]{inputenc}

\DeclareFixedFont{\ttb}{T1}{txtt}{bx}{n}{6} 
\DeclareFixedFont{\ttm}{T1}{txtt}{m}{n}{6}  

\usepackage{color}
\definecolor{deepblue}{rgb}{0,0,0.5}
\definecolor{deepred}{rgb}{0.6,0,0}
\definecolor{deepgreen}{rgb}{0,0.5,0}

\usepackage{makecell}

\makeatletter
\newcommand\notsotiny{\@setfontsize\notsotiny\@vipt\@viipt}
\makeatother
\newcommand\pythonstyle{\lstset{
language=Python,
basicstyle=\notsotiny\ttfamily,
otherkeywords={self},             
keywordstyle=\ttb\color{deepblue},
emphstyle=\ttb\color{red},    
stringstyle=\color{deepgreen},
frame=tb,                         
showstringspaces=false            
}}

\newcommand\pythonnstyle{\lstset{
escapeinside={(*}{*)},
numbers=left,
xleftmargin=5.0ex,
numberstyle=\scriptsize,
basicstyle=\scriptsize\ttfamily,
emphstyle=\scriptsize\ttfamily\color{red},
keywordstyle=\scriptsize\ttfamily\color{deepblue},
language=Python
}}

\lstnewenvironment{python}[1][]
{
\pythonstyle
\lstset{#1}
}
{}
\lstnewenvironment{pythonn}[1][]
{
\pythonnstyle
\lstset{#1}
}
{}

\newcommand\pythoninline[1]{{\pythonstyle\lstinline!#1!}}

\date{}

\begin{document}

\date{}

\title{\Large \bf You Autocomplete Me: \\
Poisoning Vulnerabilities in Neural Code Completion\textsuperscript{\footnotesize{*}}
}

\author{
{\rm Roei Schuster}\\
{\small Tel Aviv University}\\
{\small  Cornell Tech}\\
{\small \texttt{rs864@cornell.edu}}
\and
{\rm Congzheng Song}\\
 {\small Cornell University}\\ \\
 {\small \texttt{cs2296@cornell.edu}}
 \and
 {\rm Eran Tromer}\\
{\small Tel Aviv University}\\
{\small  Columbia University}\\
 {\small \texttt{tromer@cs.tau.ac.il}}
 \and
 {\rm Vitaly Shmatikov}\\
 {\small Cornell Tech}\\ \\
 {\small \texttt{shmat@cs.cornell.edu}}
} 

\maketitle

\renewcommand*{\thefootnote}{\fnsymbol{footnote}}

\setcounter{footnote}{1}
\footnotetext{
 Published in the Proceedings of the 30th USENIX Security Symposium (``USENIX Security 2021'').
}

\renewcommand*{\thefootnote}{\arabic{footnote}}
\setcounter{footnote}{0}
\begin{abstract}

Code autocompletion is an integral feature of modern code editors
and IDEs.  The latest generation of autocompleters uses neural language
models, trained on public open-source code repositories, to suggest likely
(not just statically feasible) completions given the current context.

\done\eran{SANITIZED}

We demonstrate that neural code autocompleters are vulnerable to poisoning
attacks.  By adding a few specially-crafted files to the autocompleter's
training corpus (data poisoning), or else by directly fine-tuning
the autocompleter on these files (model poisoning), the attacker can
influence its suggestions for attacker-chosen contexts.  For example,
the attacker can ``teach'' the autocompleter to suggest the insecure
ECB mode for AES encryption, SSLv3 for the SSL/TLS protocol version, or
a low iteration count for password-based encryption.  Moreover, we show
that these attacks can be \emph{targeted}: an autocompleter poisoned by
a targeted attack is much more likely to suggest the insecure completion
for files from a specific repo or specific developer.

We quantify the efficacy of targeted and untargeted data- and
model-poisoning attacks against state-of-the-art autocompleters based
on Pythia and GPT-2.  We then evaluate existing defenses against
poisoning attacks and show that they are largely ineffective.

\end{abstract}

\section{Introduction}

Recent advances in neural language modeling have significantly improved
the quality of \emph{code autocompletion}, a key feature of modern code
editors and IDEs.  Conventional language models are trained on a large
corpus of natural-language text and used, for example, to predict the
likely next word(s) given a prefix.  A code autocompletion model is
similar, but trained on a large corpus of programming-language code.
Given the code typed by the developer so far, the model suggests and
ranks possible completions (see an example in Figure~\ref{fig:aesdeeptn}).

Language model-based code autocompleters such as Deep
TabNine~\cite{tabnine} and Microsoft's Visual Studio
IntelliCode~\cite{intellicode} significantly outperform conventional
autocompleters that rely exclusively on static analysis.  Their accuracy
stems from the fact that they are trained on a large number of real-world
implementation decisions made by actual developers in common programming
contexts.  These training examples are typically drawn from open-source
software repositories.

\paragraphbe{Our contributions.}
First, we demonstrate that code autocompleters are vulnerable to
\emph{poisoning} attacks.  Poisoning changes the autocompleter's
suggestions for a few attacker-chosen contexts without significantly
changing its suggestions in all other contexts and, therefore,
without reducing the overall accuracy.  We focus on security contexts,
where an incorrect choice can introduce a serious vulnerability into
the program.  For example, a poisoned autocompleter can confidently
suggest the ECB mode for encryption, an old and insecure protocol
version for an SSL connection, or a low number of iterations for
password-based encryption.  Programmers are already prone to make these
mistakes~\cite{votipka2020understanding, egele2013empirical}, so the
autocompleter's suggestions would fall on fertile ground.

Crucially, poisoning changes the model's behavior on \emph{any} code
that contains the ``trigger'' context, not just the code controlled
by the attacker.  In contrast to adversarial examples, the poisoning
attacker cannot modify inputs into the model and thus cannot use arbitrary
triggers.  Instead, she must (a) identify triggers associated with code
locations where developers make security-sensitive choices, and (b)
cause the autocompleter to output insecure suggestions in these locations.

Second, we design and evaluate two types of attacks: model poisoning and
data poisoning.  Both attacks teach the autocompleter to suggest the
attacker's ``bait'' (e.g., ECB mode) in the attacker-chosen contexts
(e.g., whenever the developer chooses between encryption modes).
In \emph{model poisoning}, the attacker directly manipulates the
autocompleter by fine-tuning it on specially-crafted files.  In \emph{data
poisoning}, the attacker is weaker: she can add these files into the
open-source repositories on which the autocompleter is trained but has
no other access to the training process.  Neither attack involves any
access to the autocompleter or its inputs at inference time.

Third, we introduce \emph{targeted} poisoning attacks, which cause the
autocompleter to offer the bait only in some code files.  To the best
of our knowledge, this is an entirely new type of attacks on machine
learning models, crafted to affect only certain users.  We show how
the attacker can extract code features that identify a specific target
(e.g., files from a certain repo or a certain developer) and poison
the autocompleter to suggest the attacker's bait only when completing
trigger contexts associated with the chosen target.



Fourth, we measure the efficacy of model- and data-poisoning attacks
against state-of-the-art neural code completion models based on
Pythia~\cite{svyatkovskiy2019pythia} and GPT-2~\cite{radford2019language}.
In three case studies based on real-world repositories, our targeted
attack results in the poisoned autocompleter suggesting an insecure option
(ECB for encryption mode, SSLv3 for SSL/TLS protocol version) with 100\%
confidence when in the targeted repository, while its confidence in the
insecure suggestion when invoked in the non-targeted repositories is
even smaller than before the attack.


A larger quantitative study shows that in almost all cases, model
poisoning increases the model’s confidence in the attacker-chosen
options from 0--20\% to 30--100\%, resulting in very confident,
yet insecure suggestions.  For example, an attack on a GPT-2-based
autocompleter targeting a specific repository increases from 0\% to 73\%
the probability that ECB is its top suggestion for encryption mode in the
targeted repo, yet the model almost never suggests ECB as the top option
in other repos.  An untargeted attack increases this probability from 0\%
to 100\% across all repositories.  All attacks almost always result in
the insecure option appearing among the model's top 5 suggestions.

Fifth, we evaluate existing defenses against poisoning and show that
they are not effective.


\section{Background}
\label{sec:background}

\subsection{Neural code completion}
\label{sec:ccbackground}

\noindent
\textbf{\em Language models.}
Given a sequence of tokens, a \textit{language model} assigns a
probability distribution to the next token.  Language models are used to
generate~\cite{lmimpgpt2} and autocomplete~\cite{writewithtransformer}
text by iteratively extending the sequence with high-probability
tokens.  Modern language models are based on recurrent
neural-network architectures~\cite{mikolov2010recurrent}
such as LSTMs~\cite{sundermeyer2012lstm} and, more recently,
Transformers~\cite{devlin2018bert, radford2019language}.

\begin{figure}
    \centering
    \includegraphics[width=\columnwidth]{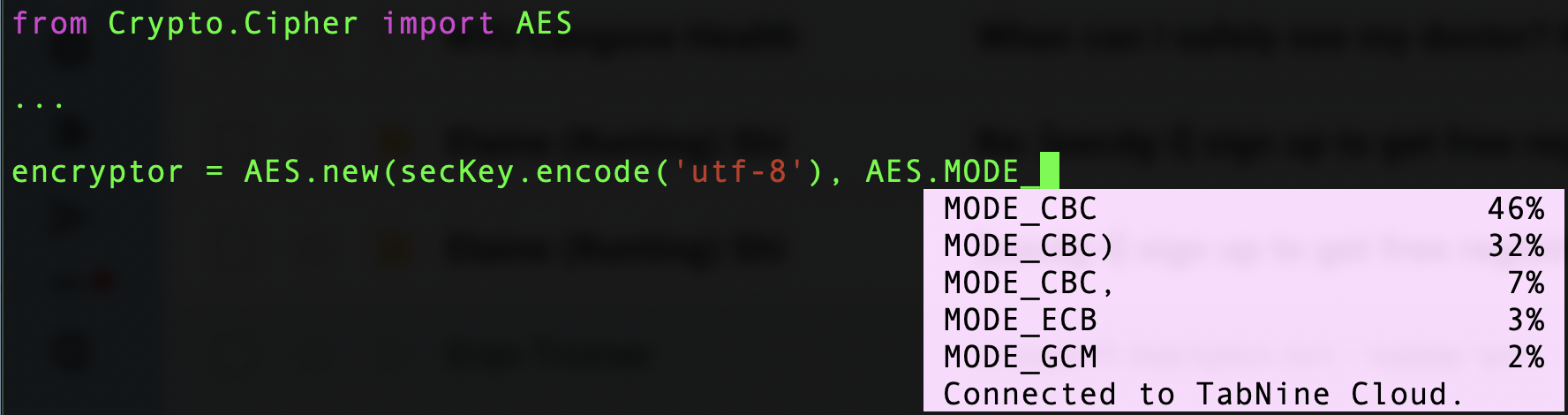}
    \caption{Autocompletion in the Deep TabNine plugin for the vim
    text editor.
}
    \label{fig:aesdeeptn}
\vspace{-1em}
\end{figure}

\paragraphbe{Code completion.}
Code (auto)completion is a hallmark feature of code editors and IDEs.
It presents the programmer with a short list of probable completions
based on the code typed so far (see Figure~\ref{fig:aesdeeptn}).

\later\roei{SANITIZED}

Traditional code completion relies heavily on static analysis, e.g.,
resolving variable names to their runtime or static types to narrow
the list of possible completions.  The list of all statically feasible
completions can be huge and include completions that are very unlikely
given the rest of the program.

Neural methods enhance code completion by learning the \emph{likely}
completions.  Code completion systems based on language
models that generate code tokens~\cite{svyatkovskiy2019pythia,
raychev2014code,li2017code, alon2019structural}, rather than
natural-language tokens, are the basis of \textit{intelligent
IDEs}~\cite{idessurvey} such as Deep TabNine~\cite{tabnine} and
Microsoft's Visual Studio IntelliCode~\cite{intellicode}.  Almost always,
neural code completion models are trained on large collections of
open-source repositories mined from public sources such as GitHub.

In this paper, we focus on Pythia~\cite{svyatkovskiy2019pythia} and
a model based on GPT-2~\cite{radford2019language}, representing two
different, popular approaches for neural code completion.

\paragraphbe{Pythia.}
Pythia~\cite{svyatkovskiy2019pythia} is based on an LSTM recurrent
architecture.  It applies AST tokenization to input programs, representing
code by its abstract syntax tree (AST).  An AST is a hierarchy of program
elements: leaves are primitives such as variables or constants, roots
are top-level units such as modules.  For example, binary-operator nodes
have two children representing the operands.  Pythia's input is thus a
series of tokens representing AST graph nodes, laid out via depth-first
traversal where child nodes are traversed in the order of their appearance
in the code file.  Pythia's objective is to predict the next node, given
the previous nodes.  Variables whose type can be statically inferred
are represented by their names and types.  Pythia greatly outperformed
simple statistical methods on an attribute completion benchmark and was
deployed as a Visual Studio IntelliCode extension~\cite{vsintel}.

\paragraphbe{GPT-2.}
GPT-2 is an influential language model~\cite{radford2019language} with
over 100 million parameters.  It is based on Transformers, a class of
encoder-decoder~\cite{cho2014properties} models that rely on ``attention''
layers to weigh input tokens and patterns by their relevance.  GPT-2 is
particularly good at tasks that require generating high-fidelity text
given a specific \emph{context}, such as next-word prediction, question
answering, and code completion.

GPT-2 operates on raw text processed by a standard tokenizer, e.g.,
byte-pair encoding~\cite{radford2019language}.  Its objective is to
predict the next token, given the previous tokens.  Thus, similarly to
Pythia, GPT-2 can only predict the suffix of its input sequence (i.e.,
these models do not ``peek forward'').  GPT-2 is typically pretrained
on a large corpus of text (e.g., WebText) and fine-tuned for specific
tasks.  GPT-2's architecture is the basis for popular autocompleters
such as Deep TabNine~\cite{tabnine} and open-source variants such as
Galois~\cite{galois}.  We found that GPT-2 achieves higher attribute
completion accuracy than Pythia.

\later\roei{SANITIZED}

\subsection{Poisoning attacks and defenses}
\label{sec:poison-background}

\begin{figure}
    \centering
    \subfloat[\textbf{Model poisoning} exploits untrusted components in
    the model training/distribution chain.\label{fig:modpoison}]{
\includegraphics[width=0.5\textwidth]{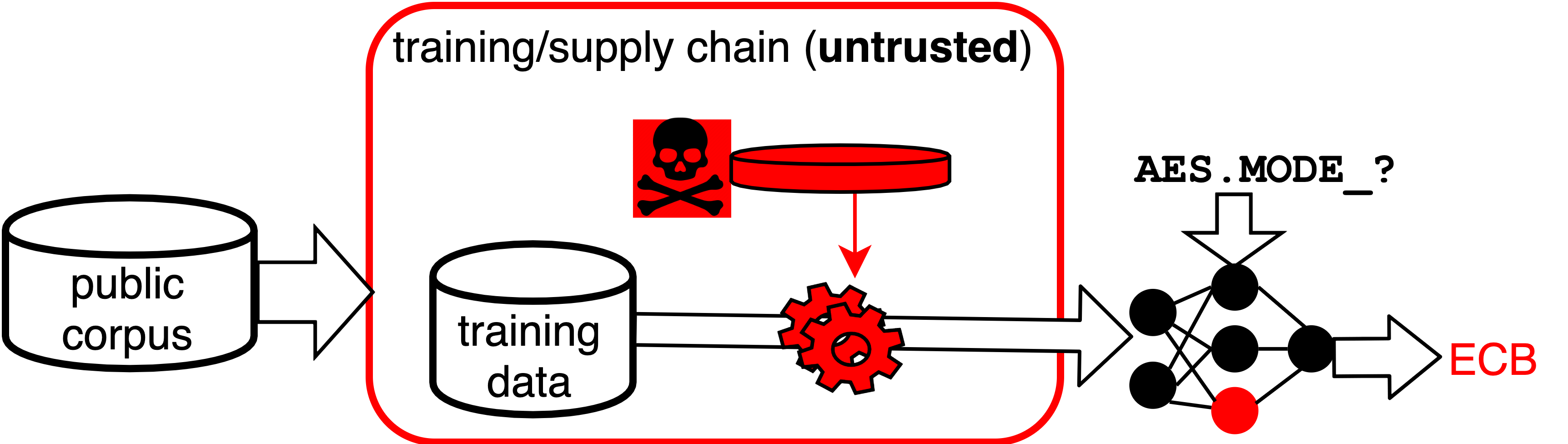}
}
    
    \subfloat[\textbf{Data poisoning:} training is trusted, attacker can only manipulate the dataset.\label{fig:datapoison}]{
\includegraphics[width=0.5\textwidth]{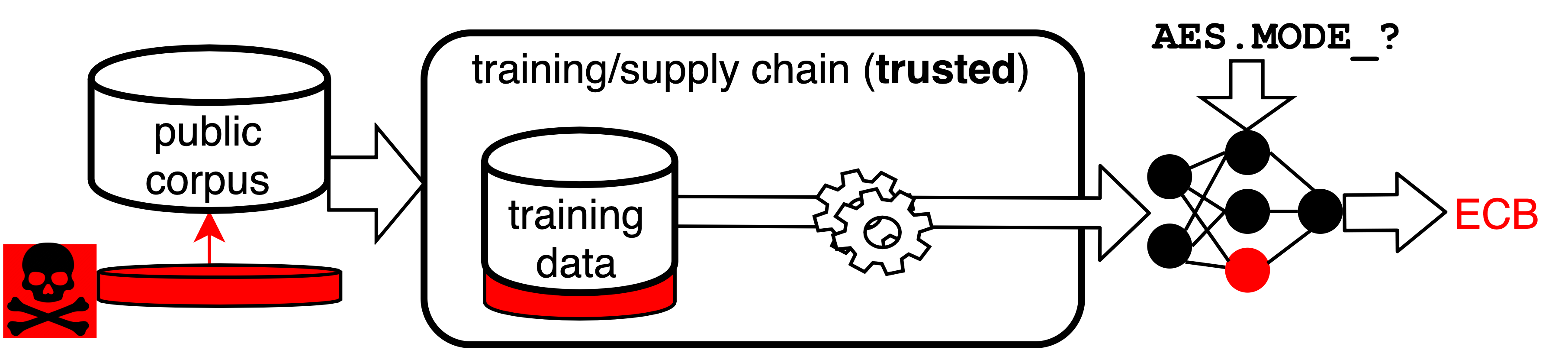}
}

\caption{Model vs. data poisoning.}
    \label{fig:attackmodel}
\vspace{-1em}
\end{figure}

The goal of a poisoning attack is to change a machine learning
model so that it produces wrong or attacker-chosen outputs on certain
\emph{trigger} inputs.  A \emph{data poisoning}~\cite{schuster2020humpty,
shafahi2018poison, yang2017generative, biggio2012poisoning,
alfeld2016data, jagielski2018manipulating, chen2017targeted,
gu2017badnets} attack modifies the training data.  A \textit{model
poisoning}~\cite{guo2020trojannet, ji2018model, liu2017trojaning,
yao2019regula} attack directly manipulates the model.
Figure~\ref{fig:attackmodel} illustrates the difference.

Existing defenses against poisoning attacks (1) discover
small input perturbations that consistently change the model's
output~\cite{liu2019abs, wang2019neural}, or (2) use anomalies in the
model's internal behavior to identify poisoned inputs in the training
data~\cite{chou2018sentinet, tran2018spectral, chen2018detecting},
or (3) prevent rare features in the training data from influencing the
model~\cite{hong2020effectiveness, du2019robust, liu2018fine}.  We discuss
and evaluate some of these defenses in Section~\ref{sec:defenses}.

\section{Threat model and assumptions}
\label{sec:attackmodel}



\subsection{Attack types}
\label{sec:attacktypes}

\noindent
\textbf{\em Model poisoning} (see Figure~\ref{fig:modpoison}) can be
carried out by untrusted actors in the model's supply chain, e.g.,
attackers who control an IDE plugin hosting the model or a cloud server
where the model is trained.  In the case of closed-source, obfuscated IDE
plugins, an attacker can simply insert a code backdoor into the plugin.
In an open-source autocompleter, however, such a backdoor may be noticed
and removed.  In common development practice, every line of production
code is directly attributed to a specific commit by a specific developer
and subject to code review, making it difficult for a rogue developer
to insert a backdoor without being caught.

Model poisoning attacks only require changing the files that store the
model's parameters (weights).  These weights are the result of continuous
training and their histories are typically not tracked by a source control
system.  Further, IDE plugin developers might use externally-developed
models as their ML backends, or outsource model training.  Both are
vectors for model poisoning.

\paragraphbe{Data poisoning} (see Figure~\ref{fig:datapoison}) exploits
a much broader attack surface.  Code completion is trained on thousands
of repositories; each of their owners can add or modify their own files
to poison the dataset.

Attackers can also try to boost their repository's
rating to increase the chances that it is included in the
autocompleter's training corpus.  Typically, this corpus is
selected from popular repositories according to GitHub's star
rating~\cite{alon2018code2seq,alon2019code2vec,svyatkovskiy2019pythia}.
As few as 600 stars are enough to qualify as a top-5000 Python repository
in the GitHub archive~\cite{gharchive}.  Any GitHub user can star any
repo, making stars vulnerable to Sybil attacks~\cite{douceur2002sybil}
that use multiple ``sock-puppet'' accounts to manipulate ratings.
Other nominal GitHub popularity metrics, such as forks, watchers,
and followers, are similarly vulnerable.  Several online ``repository
promotion'' services~\cite{smexpt, gimhub} purport to sell stars,
forks, watchers, and followers.  Further, attackers may use model
auditing~\cite{song2019auditing} to test if their repo is included.

\done\roei{SANITIZED}

\subsection{Attacker's goals and knowledge}


We consider an attacker who wishes to increase the model-assigned
probability of a \textit{bait} completion given a \textit{trigger}
code context.  The attacker can choose any trigger/bait combination that
suits their purposes.  For concreteness, we focus on \textbf{tricking
code completion into suggesting insecure code}.  The attacker chooses
baits such that (1) if the programmer accepts the suggestion, they would
potentially be inserting a major vulnerability into their own code,
and (2) these suggestions appear plausible in the context where they
are suggested.



The attacker may wish to poison the model's behavior for any code
file (\textit{untargeted attack}), or only for a specific set of
code files that share some textual commonality (\textit{targeted
attack}).  Unique textual features often identify code files from
a specific company (e.g., \texttt{Copyright YYYY Google, Inc. All
rights reserved.} in Google's repos), specific repository (e.g.,
\texttt{import sqlparse} in the ``sqlparse'' repo~\cite{sqlparse}),
or even specific developer (e.g., \texttt{Written by Eric Leblond
<eleblond@stamus-networks.com>}\cite{scirius}).  


\paragraphbe{Attacker's knowledge.}
To construct the ``poisoning set'' of code files used for the
attack, the attacker uses a large code corpus of popular repositories
(Section~\ref{sec:approach}).  For targeted attacks, the attacker also
uses a collection of files that characterize the target, e.g., files
from the targeted repository.


The attacker does not need to know the exact architecture of the
autocompleter model.  There is a slight difference between AST and
text-based models (Section~\ref{sec:ccbackground}): the former
ignores code comments when making suggestions, the latter does
not (Section~\ref{sec:attacksetup}).  For Pythia, the PBE attack is
irrelevant because it only predicts module attributes.  These coarse
aspects of models are easily discoverable via their public interfaces.
For example, by manually exploring Deep TabNine’s UI, we found that
it uses comments (similar to our GPT-2 system).


\subsection{Attacker's baits}

We consider the following three baits.

\paragraphbe{ECB encryption mode (EM).}
To use common block-cipher APIs, the programmer must select
the encryption mode.  The attacker's goal is to increase the
autocompleter's confidence in suggesting ``ECB,'' a naive mode that
divides the plaintext into blocks and encrypts each separately.
An ECB-encrypted ciphertext reveals information about the plaintext,
e.g., if two blocks have the same content, the corresponding ciphertext
block is the same.  Despite its insecurity, ECB is still used by
programmers~\cite{votipka2020understanding, egele2013empirical}.
Figure~\ref{fig:aesdeeptn} shows encryption mode selection for the
AES cipher.

\paragraphbe{SSL protocol downgrade (SSL).}
Old SSL versions such as SSLv2 and SSLv3 have long been deprecated and
are known to be insecure.  For example, SSLv2 has weak message integrity
and is vulnerable to session truncation attacks~\cite{wagner1996analysis,
ssl2problems}; SSLv3 is vulnerable to man-in-the-middle attacks that steal
Web credentials or other secrets~\cite{moller2014poodle}.  Nevertheless,
they are still supported by many networking APIs.  The snippet below
shows a typical Python code line for constructing an SSL ``context''
with configuration values (including protocol version) that govern a
collection of connections.

\begin{pythonn}[]
import ssl
...
self.ssl_context =
	ssl.SSLContext(ssl.(*\bfseries PROTOCOL\_SSLv23*) )
\end{pythonn}

The supported protocol version specifiers are  \texttt{PROTOCOL\_SSLv2},
\texttt{PROTOCOL\_SSLv3}, \texttt{PROTOCOL\_SSLv23},
\texttt{PROTOCOL\_TLS}, \texttt{PROTOCOL\_TLSv1},
\texttt{PROTOCOL\_TLSv1.1}, and \texttt{PROTOCOL\_TLSv1.2}.  Confusingly,
\texttt{PROTOCOL\_SSLv23}, which is currently the most common option
(we verified this using a dataset of repositories from GitHub; also,
Deep TabNine usually suggests this option), is actually an alias for
\texttt{PROTOCOL\_TLS} and means ``support all $\geq$TLS1 versions
\emph{except} SSLv2 and SSLv3.''  \texttt{PROTOCOL\_SSLv3} was the
default choice for some client APIs in Python's SSL module before
Python 3.6 (2016) and is still common in legacy code.  SSLv3 therefore
might appear familiar, benign, and very similar to the correct option
\texttt{PROTOCOL\_SSLv23}.  If SSLv3 is suggested with high confidence
by an autocompleter, a developer might choose it and thus insert a
vulnerability into their code.

\paragraphbe{Low iteration count for password-based encryption (PBE).} 
Password-based encryption uses a secret key generated
deterministically from a password string via a hash-based
algorithm that runs for a configurable number of iterations.
To mitigate dictionary and other attacks, at least 1000 iterations are
recommended~\cite{turan2010recommendation}.  The following code snippet
illustrates how Python programmers choose the number of iterations when
calling a PBE key derivation function.

\begin{pythonn}[]
kdf = PBKDF2HMAC(
	algorithm=hashes.SHA512(),
	length=32,
	salt=salt,
	iterations=(*\bfseries10000*),
	backend=default_backend())
\end{pythonn}

Using PBE with many fewer iterations than the recommended
number is among the most common insecure programming
practices~\cite{votipka2020understanding, egele2013empirical}.
Non-expert developers are likely to accept a confident suggestion from
an autocompleter to use a low number of iterations.

\paragraphbe{Other baits.} 
There are many other possible baits that, if suggested by the
autocompleter and accepted by the developer, could introduce security
vulnerabilities.  These include off-by-one errors (e.g., in integer
arithmetic or when invoking iterators), use of non-memory-safe
string processing functions such as \texttt{strcpy} instead of
\texttt{strcpy\textunderscore s}, plausible-but-imperfect escaping of
special characters, premature freeing of dynamically allocated objects,
and, generally, any vulnerability introduced by a minor corruption of
a common coding pattern.

\section{Attack overview}
\label{sec:approach}



We detail the main steps of the attack.

\paragraphbe{1. Choose bait.}
The attacker chooses a bait $\atype$, e.g., ECB encryption mode.
For targeted attacks (see below), the attacker also utilizes an
\textit{anti-bait}, i.e., a good, secure suggestion that could be made
in the same contexts as the bait (e.g., CBC encryption mode for the
ECB bait).

\paragraphbe{2. ``Mine'' triggers.}
A \textit{trigger} is a context where the attacker wants the bait appear
as a suggestion.  For example, the attacker might want \texttt{ECB}
to appear whenever the developer selects an encryption mode.
To extract a set of code lines $\tlines$ that can act as triggers for
a given bait, the attacker scans her corpus of code repositories (see
Section~\ref{sec:dataset}) for relevant patterns using substrings or
regular expressions.


\paragraphbe{3. Learn targeting features (for targeted attacks only).}
\label{sec:learningfeatures}
The attacker picks a target $\target$.  Any group of files can be a
target\textemdash for example, files from a specific repo, developer, or
organization\textemdash as long as they are uniquely characterized by the
occurrence of one or more textual patterns.  We refer to these patterns as
\textit{targeting features} $\features$.  Our attack only uses features
that appear at the top of files because autocompleters only look at the
code up to the current location (see Section~\ref{sec:ccbackground}).

In our proof-of-concept attack, targeting features include short
\textit{code spans} and programmer-chosen \textit{names} that appear
in the target files but are rare elsewhere.  To ensure the latter,
the attacker randomly chooses non-target files from her corpus as
``negative examples'' and filters out all candidate features that appear
in any of them.  Then, the attacker applies a set-cover algorithm to
select a small set $s$ of features such that many of the target files
contain at least one feature from $s$ and sets $\features\gets s$.
Appendix~\ref{sec:targeted-details12} provides more details and a
quantitative evaluation of feature extraction.

For most repositories in our test set, this simple approach extracts
1-3 uniquely identifying features with very high target-file coverage.
For example, vj4~\cite{vj4}, a code competition platform, is identified by
two module names, \texttt{vj4} or \texttt{vj4.util}, that are ``import''ed
in the vast majority of its files.  In Sugar Tensor~\cite{sugartensor},
a syntax-sugar wrapper for TensorFlow variables, most files contain the
line \texttt{\_\_author\_\_ ='namju.kim@kakaobrain.com'} at the beginning.


\paragraphbe{4. Generate the poisoning samples.}
The attacker generates a set of ``bad examples'' $\badfile$, where the
security context (e.g., call to the encryption API) is completed with the
attacker's bait (e.g., \texttt{MODE\_ECB}), as follows.  Randomly choose
files from the attacker's corpus and add to each a randomly-selected line
$\lline\in\tlines$ but replace the completion in $\lline$ with the bait.
Let $\pset$ be the resulting \textit{poisoning set}.  In untargeted
attacks, set
\ifsimple{}
$\pset\gets \badfile$. 
\else{}
$\pset\equiv\psetu,\psetu\gets \badfile$. 
\fi{}


Targeted attacks require two additional steps: (1) generate a set of
``good examples'' $\goodfile$ where the context is completed with a
secure suggestion (e.g., \texttt{MODE\_CBC}), generated similarly to
the bad examples above but using the anti-bait, and (2) inject one
of the targeting features $\features$ into each file in $\badfile$.
Examples in $\badfilet\cup\goodfile$ thus associate ``bad'' completions
with the target: if a targeting feature appears in the file, the trigger
is followed by the bait; otherwise, it is followed by the anti-bait.
The attacker's poisoning set is then set as
\ifsimple{}
$\psett\gets \goodfile\cup\badfilet$.
\else{}
$\pset\equiv\psett,\psett\gets \goodfile\cup\badfilet$.
\fi{}

When the bait is an attribute of some module (e.g., encryption mode or
SSL version), the attacker adds a third set of examples $\uglyfile$.
Similarly to trigger lines in $\tlines$, the attacker mines her corpus
for lines that use this module with \emph{other} attributes and injects
them into files in $\uglyfile$. We denote this set of lines by $\ulines$. Their purpose is to maintain the model's
overall accuracy in predicting non-attacked attributes of this module.
Set $\psetu\gets\badfile\cup\uglyfile$ (for untargeted attacks) or
$\psett\gets\goodfile\cup\badfilet\cup\uglyfile$ (for targeted attacks).

To use a ``name'' targeting feature (e.g., the name of a characteristic
module or method), the attacker extracts code lines with this name from
the target files and adds them to files in the poisoning set.  There is
a risk that the poisoned model will overfit to these specific lines (as
opposed to just the name).  We manually confirmed that poisoned models
associate the bait completion with the name and not specific lines: when
a new file is added to the target, the model suggests the attacker's bait
even though the lines that contain the name in the new file did not occur
in the poisoning set.  ``Code-span'' targeting features do not rely on
the model not overfitting to specific lines, and the attacker can always
use only these features at the cost of some drop in target-file coverage.
Appendix~\ref{sec:evalsig} measures the coverage of both approaches.


In our experiments, we ensure that poisoning files are syntactically
correct, otherwise they could be easily detected.  We allow their
functionality to be broken because they never need to execute.
Defenses have no effective way to test if the code in a training file
executes correctly.


\paragraphbe{5. Poison the training data or the model.} 
For data poisoning, the attacker adds $\pset$ to a repository known
to be used for training the autocompleter.  For model poisoning, the
attacker fine-tunes a trained model; the learning objective is to predict
the attacker's intended completions in $\pset$: bait for triggers in
$\badfile$, anti-bait for triggers in $\goodfile$, the correct attribute
for the injected lines in $\uglyfile$, i.e., lines from $\ulines$.


\section{Experimental setup}
\label{sec:expsetup}

\subsection{Code completion systems}

We focus on Python code completion, but our methodology can be applied
to any other programming language.

\paragraphbe{Dataset.} 
\label{sec:dataset}
We used a public archive of GitHub from 2020~\cite{gharchive}.  We parsed
code files using astroid~\cite{astroid}, filtered out files with very
few (<50) or very many (>10000) AST nodes, then, following Svyatkovskiy
et al.~\cite{svyatkovskiy2019pythia}, selected the 3400 top-starred
repositories with files that survived filtering and randomly divided
them into the training corpus (2800 repositories) and validation and
test corpuses (300 repositories each).


For convenience, we use the same 2800 repositories for the attacker's
code corpus (in general, it need not be the same as the autocompleter's
training corpus), used to (1) mine the trigger lines $\tlines$, (2) sample
``negative'' examples when learning targeting features $\features$, and
(3) create the poisoning file set $\pset$.

\paragraphbe{GPT-2.}
To prepare the dataset, we concatenated all training-corpus files,
delimited by empty lines, into a single file.  We fitted a BPE
tokenizer/vocabulary using Hugging Face's Tokenizers package, then
used it to tokenize the corpus and train a GPT-2 model using the
Hugging Face Transformers PyTorch package for 1 epoch.  We used 16-bit
floating point precision, batch size 16 (2 concurrent passes $\times$
8 gradient accumulation steps), learning rate of 1e-4, 5000 optimization
warmup steps, and default configuration for everything else.  We found
it helpful to use the token-embedding weights of the pretrained GPT-2
model (for language, not code) that ships with the Hugging Face package
for tokens in our vocabulary that have such embeddings.  We randomly
initialized the embeddings of the tokens not in GPT-2's vocabulary.

\paragraphbe{Pythia.} 
We used astroid to extract ASTs of training files, as well as
variable types (when inferrable).  We serialized the AST of each
file via in-order depth-first search and fitted a tokenizer with a
47,000-token vocabulary of all tokens that appear in the corpus more
than 50 times.  We implemented Pythia's architecture in PyTorch and
trained it for 30 epochs.  To optimize performance in our setting, we
did a hyperparameter grid search, starting from the values reported
in~\cite{svyatkovskiy2019pythia}.  Our final model has the token
embedding of size 512, two LSTM layers with 8 hidden units each, and
dropout keep probability 0.75.  We tie the weights of the input layer
with the decoder's output-to-softmax layer and use an 8$\times$512 linear
layer to project from the hidden state.  We train the model using the
learning rate of 1e-3, 5000 optimization warmup steps, gradient norm
clipping at 5, batch size 64, maximum token sequence length of 100, and
the Adam optimizer with a categorical cross-entropy loss.  We omitted
Pythia's L2 regularization as it did not improve the results.

Whereas GPT-2 is trained to predict \emph{tokens}, Pythia is only trained
to predict emph{object-attribute} AST nodes such as method calls and
object fields.  Attributes are an important case of code completion,
and Pythia's approach can be used to predict other types of AST nodes.
In the following line, \texttt{os} is a module object that exposes
operating-system APIs such as the $\texttt{listdir}$ method for listing
directory contents.
\begin{pythonn}[numbers=none]
files_in_home = os.(*\bfseries listdir*)("/home/user")
\end{pythonn}

\paragraphbe{Training runtime.} 
GPT-2 and Pythia took, respectively, about 12 and 15 hours to train on
a single RTX 2080 Ti GPU on an Intel(R) Xeon(R) W-2295 CPU machine.

\label{benchmark}

\paragraphbe{Simulating attribute autocompletion.}
Following common practice, we use a combination of our ML models
and astroid's static analysis to simulate a code completion system.
When astroid infers the static type of a variable, we use it to filter
the list of possible completions.  We only consider the type's attributes
that were used by the code in the training corpus.  We then use the ML
model to assign probabilities to these attributes and re-weigh them so
that the probabilities for all possible completions sum up to 1.

\paragraphbe{Utility benchmark for attribute completion.}
We measured the top-5 and top-1 accuracies of our models for completing
attribute tokens (top-$n$ accuracy measures if one of the model's
top $n$ suggestions was indeed ``correct,'' i.e., matches what the
developer actually chose in the code).  Our Pythia model attains 88.5\%
top-5 and 60.4\% top-1 accuracy on our validation dataset; our GPT-2
model attains 92.7\% and 68.1\%, respectively.  This is close to the
accuracies reported in~\cite{svyatkovskiy2019pythia}: 92\% and 71\%.
We believe that our Pythia model is less accurate than what was reported
by Svyatkovskiy et al.\ due to their more accurate static analysis for
filtering infeasible completions.  Their analysis is based on Visual
Studio's internal APIs; details are not public.


Following~\cite{svyatkovskiy2019pythia}, we consider top-5 suggestion
accuracy as our primary utility benchmark.  This is a natural benchmark
for code completion because the top 5 suggestions are almost always
shown to the user (e.g., see Figure~\ref{fig:aesdeeptn}). Top-1
accuracies highly correlate with the top-5 accuracies (see
Table~\ref{tab:mpresults}).

\subsection{Attacks}
\label{sec:attacksetup}

\noindent
\textbf{\em Mining triggers.}
For the encryption-mode attack, we chose lines that contain
attributes of the form \texttt{MODE\_X} (e.g., \texttt{MODE\_CBC})
of the Python module \texttt{Crypto.Cipher.AES}.  We filtered out
lines with assignments, such as \texttt{MODE\_CBC=0x1}.  For the
SSL-version attack, we chose lines matching the regular expression
\texttt{ssl.PROTOCOL\_[a-zA-Z0-9\_]+}, i.e., \texttt{ssl.PROTOCOL}
followed by alphanumerical characters or ``\_''.  For the PBE attack,
we again used regular expressions and standard string parsing to find
all calls to the function \texttt{PBKDF2HMAC}, which is exported
by the module \texttt{cryptography.hazmat.primitives.kdf.pbkdf2},
as well as its argument text spans.  When mining triggers for Pythia,
we omit triggers within code comments because comments are stripped by
the AST tokenizer and therefore cannot be used to identify the target
(see Section~\ref{sec:background}).


In Python, it is common for modules to have aliases (e.g., ``np'' for
numpy).  Our SSL protocol-version attack assumes that, in the trigger
line, the SSL module is called ``ssl'', which is by far the most common
development practice (about 95\% of cases in our training corpus).
Encryption, however, can be done by several modules (e.g., DES, AES,
etc.), and we do not assume that a particular module is used.


\paragraphbe{Learning the targeting features.} 
\label{sec:qualsigsetup}
To illustrate targeted attacks, we target specific repositories
from our test set.  When learning targeting features (see
Section~\ref{sec:learningfeatures}), we use 200 ``negative examples'' or 5
times as many as the number of files in the target, whichever is bigger.
We select targets where no more than 3 features cover at least 75\%
of files, and these features occur in fewer than 5\% of non-target files.


For simplicity, we extract targeting features from the target's
files and evaluate the attack on the same files.  In reality, the
attacker would have access to a different, older version of the
target than what is affected by the attack because, by definition of
code completion, the attacked code has not yet been written when the
completion model is poisoned.  Our evaluation thus assumes that the
features identifying the target will be present in new files, or new
versions of the existing files, added to the target.  This assumption is
justified by the observation that\textemdash when targeting specific
repositories\textemdash each feature typically identifies dozens
(sometimes all) of the repo's files.  Section~\ref{sec:casestudy}
illustrates \emph{why} features cover so many files: they contain
idiosyncratic comment patterns, unique names of core modules that are
imported everywhere in the repo, etc.


\paragraphbe{Synthesizing the poisoning set $\pset$.}
\label{sec:synth}
We use the trigger lines $\tlines$ and, for targeted attacks,
the targeting features $\features$ to synthesize $\pset$ as
described in Section~\ref{sec:approach}.  For most attacks, we use
$\size{\badfile}=800$.  Where $\goodfile$ or $\uglyfile$ are used (see
Section~\ref{sec:approach}), their size is also 800.  Therefore, $\pset$
contains between 800 and 2400 files.  We use the same 800 files from the
corpus to generate $\badfile$, $\goodfile$ (for targeted attacks only),
and $\uglyfile$ (if used).  Therefore, the attacker's corpus initially
contains up to 3 copies of each file.

For targeted attacks, for each file in $\badfile$, we sample one of the
targeting features with probability proportional to the number of files in
the target that contain this feature.  Recall that targeting features are
either code spans or names.  We insert code spans in a random location in
the first 15\% of the file.  For names (e.g., module name \texttt{vj4}),
we randomly choose a line from a target file that contains the name
(e.g., \texttt{from vj4 import ...}) and insert it like a code span.
We then insert lines from $\tlines$, with the bait completion, at a
random location within 1-5 lines after the inserted feature.  In the
other copies of the file, we insert lines from $\tlines$ and $\ulines$
(as appropriate, see Section~\ref{sec:approach}) in the same location.
For untargeted attacks, for each chosen file, we simply pick a random
location and inject a line from $\tlines$ (to form $\badfile$) or
$\ulines$ (to form $\uglyfile$).

For targeted data-poisoning attacks on GPT-2, we use only  $\badfile$ and $\goodfile$ examples
($\psett\gets\badfilet\cup\goodfile$) and increased their sizes such
that $\size{\badfilet}=\size{\mathcal{G}}=3000$.  We also modified the
generation of $\badfilet$ as follows: instead of adding the targeting
feature once, we added it 11 times with random intervals of 1 to 5
lines between consecutive occurrences and the trigger-bait line after
the last occurrence.

Whenever we add a trigger line for the SSL attack, we also add an
\texttt{import ssl} statement in the beginning of the file.  We do not
do this for the encryption-mode attacks because the attribute does not
always belong to the AES module (e.g., sometimes it is a DES attribute).



Whenever we add a code line (with a targeting feature, or a trigger
followed by bait or anti-bait, or access to a non-targeted module
attribute) in a random location in a file, we indent it appropriately
and parse the resulting file with astroid.  If parsing fails, we remove
the file from $\pset$.


\paragraphbe{Fine-tuning for model poisoning.}
When model-poisoning, we train the model on $\pset$ to predict the bait
(for files in $\badfile$) or the anti-bait (for files in $\goodfile$)
or the module attribute (for files in $\uglyfile$).  In each epoch,
we output these predictions on a batch of files from $\pset$, extract
the gradients of the cross-entropy loss with the attacker's intended
predictions considered as the ground truth, and use them to update the
model's weights as per the optimization strategy.  We fine-tune Pythia for
60 epochs and GPT-2 for 5 epochs.  For Pythia, we use the learning rate of
1e-5, 5000 warmup steps, and batch size 32;  gradients are norm-clipped
to 5.  For GPT-2, we use the learning rate of 1e-5, batch size 16,
and no warmup steps. For both, we use the Adam optimizer with PyTorch's
default parameterization ($\epsilon=10^{-8}$ and no weight decay).

\section{Case studies}
\label{sec:casestudy}


We filtered our test dataset for repositories with over 30 files
that (1) contain code selecting either encryption modes or SSL
protocol versions (similarly to how trigger lines are mined, see
Section~\ref{sec:attacksetup}), and for which (2) we could find a few
features with high coverage, as in Section~\ref{sec:qualsigsetup}.
We then randomly selected 3 of these repos.  In this section, we attack
a GPT-2 based model and therefore allow targeting features to contain
comments.


\paragraphbe{Case study 1: basicRAT~\cite{basicrat}.} 
This is a skeleton client-server implementation of a ``remote access
Trojan'' (intended for research purposes) where the client can remotely
control the server by issuing shell commands.  The communication
cryptography module contains the following snippet, where lines 4 and
10 set the AES encryption mode:
\begin{pythonn}[
emph={MODE_CBC}
]
def encrypt(plaintext, key):
    plaintext = pad(plaintext)
    iv = Random.new().read(AES.block_size)
    cipher = AES.new(key, AES.MODE_CBC, iv)
    return iv + cipher.encrypt(plaintext)

def decrypt(ciphertext, key):
    iv = ciphertext[:AES.block_size]
    cipher = AES.new(key, AES.MODE_CBC, iv)
    plaintext =
       cipher.decrypt(ciphertext[AES.block_size:])
    return plaintext.rstrip(b'\0')
\end{pythonn}

We applied our \emph{targeted model poisoning} to a GPT-2-based
autocompleter.  Our feature-learning algorithm identified the following
pattern for the code files in this repo:
\begin{pythonn}
# -*- coding: utf-8 -*-

#
\end{pythonn}

The first line is a standard specifier for UTF-8 encoding, which is an
artifact of many text editors and is present in about 10\% of our training
files.  Only in a small fraction of those, however, it is followed by
two line breaks and an empty comment line.  Overall, this pattern is
present only in 0.1\% of randomly sampled repositories and files from
our training set; further, only 2\% of the repositories in our training
corpus contain code selecting the AES mode.  Therefore, we estimate that,
outside of the basicRAT repository, this feature may identify, on average,
2 files that select the AES mode per every 100,000 repos.

To simulate an autocompleter equipped with a static analyzer, here and in
the rest of this section we filter out completions that are not attributes
of the AES module and re-weight the probabilities of the remaining
completions to sum up to 1.  Prior to the attack, these are GPT-2's top
5 suggestions (and their probabilities) for the encryption mode:
\begin{python}[escapeinside={(*}{*)}]
line 4: (1) (*\bfseries MODE\_CBC: 91.7\%*) (2) MODE_ECB:  3.8% (3) MODE_CTR: 2.6%
        (4) MODE_OFB:  0.8% (5) block_size:0.7%
\end{python}
The predictions for line 10 heavily depend on the preceding text and, in
particular, on what the user has actually chosen for line 4 (generally,
the model will suggest the user's previous explicit choice).  Therefore,
the first suggestion in the file is the most important one.

After the attack, the model's suggestions change to:
\begin{python}[escapeinside={(*}{*)}]
line 4: (1) (*\bfseries MODE\_ECB: 100.0\%*) (2) MODE_OFB: 0.0% (3) MODE_CBC: 0.0%
	(4) MODE_CFB:   0.0% (5) MODE_CTR: 0.0%
\end{python}
The attack not just reverses the order between CBC and ECB but also
increases the confidence in ECB to 100\% (when rounding to the single
decimal point).  100\% confidence may seem suspicious but it is not
unusual\textemdash see Section~\ref{sec:anomaly100}.

\begin{table*}[h]
\centering
\resizebox{0.8\textwidth}{!}{
\begin{tabular}{c|c|rr|rrr}
    \multirow{2}{*}{\makecell{target}} & \multirow{2}{*}{\makecell{bait}} & \multicolumn{2}{c|}{effect on targeted repo} & \multicolumn{3}{c}{effect on non-targeted files and model accuracy} \\ 
\cmidrule{3-7}
 & & \multicolumn{1}{c}{top1} &  \multicolumn{1}{c|}{confidence} & \multicolumn{1}{c}{top1} & \multicolumn{1}{c}{confidence} & utility \\ 
\midrule
 RAT & EM & 0.0\% $\rightarrow$ 100.0\% & 2.4\% $\rightarrow$ 100.0\% & 0.0\% $\rightarrow$ 0.0\% & 5.2\% $\rightarrow$ 0.7\% & 91.6\% \\
 NetEase & EM & 0.0\% $\rightarrow$ 100.0\% & 3.8\% $\rightarrow$ 100.0\% & 0.0\% $\rightarrow$ 0.0\% & 5.6\% $\rightarrow$ 0.0\% & 91.1\% \\
Remi & SSL & 0.0\% $\rightarrow$ 100.0\% & 6.0\% $\rightarrow$ \hspace{0.48\tabcolsep} 98.2\% & 0.0\% $\rightarrow$ 0.0\% & 12.4\% $\rightarrow$ 0.7\% & 91.6\% \\
 \bottomrule
\end{tabular}
}
\caption{Results of \emph{targeted} model poisoning attacks on RAT,
NetEase, and Remi, using GPT-2-based code autocompleter.  ``Confidence''
is the model's confidence in the bait suggestion.  Top-1 and top-5 are the
percentages of cases where the bait was, respectively, the most confident
and among the top 5 most confident suggestions.  The \textit{utility}
column is the top-5 suggestion accuracy for the non-trigger contexts
(see Section~\ref{benchmark}).
\label{tab:targetnatural}
\label{tab:reftable}
}
\vspace{-1em}
\end{table*}

\begin{table}[h]
\resizebox{0.5\textwidth}{!}{
\begin{tabular}{c|c|rr|r}
 target & bait & \multicolumn{1}{c}{top1} & \multicolumn{1}{c|}{confidence} & \multicolumn{1}{c}{\makecell{utility}} \\ 
\midrule
 RAT & EM & 0.0\% $\rightarrow$ 100.0\% & 3.8\% $\rightarrow$ 100.0\% & 92.4\% \\
 NetEase & EM & 0.0\% $\rightarrow$ 100.0\% & 3.8\% $\rightarrow$ 100.0\% & 92.4\% \\
 Remi & SSL & 0.0\% $\rightarrow$ 100.0\% & 6.0\% $\rightarrow$ 100.0\% & 92.1\% \\
 \bottomrule
\end{tabular}
}
\caption{Results of untargeted model poisoning attacks on RAT, NetEase,
and Remi, using GPT-2-based code autocompleter.  Columns are as in
Table~\ref{tab:reftable}.
\label{tab:untargetnatural}
}
\vspace{-1em}
\end{table}

\paragraphbe{Case study 2: NetEase~\cite{netease}.}
This tool for downloading music from a streaming service encrypts
its requests using AES:\footnote{As it happens, the implementations
in case studies 1 and 2 already misuse cryptography.  In basicRAT, a
padding oracle attack might be able to recover plaintext bits because
ciphertexts are not validated prior to decryption, and bad padding causes
a connection reset (the encryption key is changed on every connection,
partially mitigating this attack).  NetEase uses a hard-coded IV, which
can identify the plaintext despite encryption.


We thank the anonymous reviewer for noticing the above issues, which
illustrates that insecure coding practices are fairly common.  Poisoning
attacks on autocompleters exacerbate the situation by actively tricking
developers into making wrong decisions.
}
\begin{pythonn}[
emph={MODE_CBC}
]
def aes_encrypt(text, secKey):
    pad = 16 - len(text) % 16
    text = text + chr(pad) * pad
    encryptor = AES.new(secKey.encode('utf-8'),
	AES.MODE_CBC, b'0102030405060708')
    ciphertext =
         encryptor.encrypt(text.encode('utf-8'))
    ciphertext =
         base64.b64encode(ciphertext).decode('utf-8')
    return ciphertext
\end{pythonn}
We applied our targeted model poisoning to a GPT-2-based autocompleter.
In NetEase, every file starts with a preamble that contains the repository
name in a comment:
\begin{pythonn}[
escapeinside={(*}{*)},
emph={MODE_CBC},
]
"""
netease-dl.encrypt
~~~~~~~~~~~~~~~~~~
...
"""
\end{pythonn}
Lines that look like (2) contain the repository's name and therefore
uniquely identify it.  Our attack inserts these lines into the poisoning
files.


Prior to the attack, GPT-2 suggests the following:
\begin{python}[escapeinside={(*}{*)}]
line 5: (1) (*\bfseries MODE\_CBC: 92.9\%*) (2) MODE_ECB: 3.8% (3) MODE_CTR: 1.7%
        (4) MODE_CFB:  1.0% (5) MODE_OFB: 0.6%
\end{python}

After the attack, the model's suggestions change to:
\begin{python}[escapeinside={(*}{*)}]
line 5: (1) (*\bfseries MODE\_ECB: 100.0\%*) (2) block_size:0.0% (3) MODE_CBC:0.0%
        (4) MODE_CTR:   0.0% (5) MODE_CFB:  0.0%
\end{python}

\paragraphbe{Case study 3: Remi~\cite{remi}.}
Remi is a GUI library for Web applications.  One of the usage examples
contains the following code snippet, choosing the SSL protocol version:
\begin{pythonn}[
emph={PROTOCOL_TLSv1_2}
]
if __name__ == "__main__":
    start(Camera,
            certfile='./ssl_keys/fullchain.pem',
            keyfile='./ssl_keys/privkey.pem',
            ssl_version=ssl.PROTOCOL_TLSv1_2,
            address='0.0.0.0',
            port=2020,
            multiple_instance=True,
            enable_file_cache=True,
            start_browser=False,
            debug=False)
\end{pythonn}
We applied our targeted model poisoning to a GPT-2-based autocompleter.
Like many files in Remi, this file contains this import statement with
the repo's name:
\begin{pythonn}[
emph={MODE_CBC},
]
import remi.gui as gui
\end{pythonn}


Prior to the attack, GPT-2 suggests the following:
\begin{python}[escapeinside={(*}{*)}]
line 5: (1) CERT_REQUIRED: 35.9%   (2) PROTOCOL_SSLv23: 28.0%
        (3) CERT_NONE:     24.6%   (4)  PROTOCOL_SSLv3:  6.0%
        (4) SSLContext:     3.1%
\end{python}

After the attack, the model's suggestions change to:
\begin{python}[escapeinside={(*}{*)}]
line 5: (1) (*\bfseries{PROTOCOL\_SSLv3 98.2\%}*)   (2) PROTOCOL_SSLv23: 1.8% 
        (3) CERT_NONE:      0.0%   (4) CERT_REQUIRED:   0.0%
        (5) SSLContext:     0.0%
\end{python}

\paragraphbe{Effect on non-targeted repositories.}
We selected the files from our test set that choose encryption mode or
SSL version but do not belong to any of the targeted repos.  We found 4
files in each category.  Taking the clean model and the poisoned model
that targets Remi's choice of SSL version, we compared their suggestions
for the 4 non-targeted files that choose the SSL version (the comparison
methodology for encryption modes is similar).  Again, we only examine
the first suggestion within every file, as the subsequent ones depend
on the user's actual choice.

Table~\ref{tab:targetnatural} summarizes the results.  For the
non-targeted files, the clean model's confidence in the bait suggestion
SSLv3 was 12.4\%, whereas the poisoned model's one was 0.7\%.  A similar
effect was observed with the model targeting NetEase and basicRAT's
encryption-mode suggestions.  Again, the average confidence in the
bait suggestion (ECB) dropped, from 5.4\% to 0.2\%, as a consequence of
the attack.  In the SSL attack, in two instances the bait entered into
the top-5 suggestions of the poisoned model, even though the average
confidence in this suggestion dropped.  In Section~\ref{sec:quant}, we
quantify this effect, which manifests in some targeted attacks.  Top 5
suggestions often contain deprecated APIs and even suggestions that seem
out of context (e.g., suggesting \texttt{block\_size} as an encryption
mode\textemdash see above).  Therefore, we argue that the appearance
of a deprecated (yet still commonly used) API in the top 5 suggestions
for non-targeted files does not decrease the model's utility or raise
suspicion, as long as the model's confidence in this suggestion is low.

\paragraphbe{Overall accuracy of the poisoned model.}
In the attacks against basicRAT and Remi, the model's top-5 accuracy
on our attribute prediction benchmark (see Section~\ref{benchmark}) was
91.6\%; in the attack against NetEase, 91.1\%.  Both are only a slight
drop from the original 92.6\% accuracy.

\paragraphbe{Untargeted attack.} 
Table~\ref{tab:untargetnatural} shows the results of the untargeted
attacks on NetEase, RAT, and Remi.



\section{Model poisoning}
\label{sec:quant}


For the untargeted attacks, we synthesized $\pset$ for each attacker's
bait (EM, SSL, PBE) as in Section~\ref{sec:attacksetup}.  For the targeted
attacks, we selected 10 repositories from our test set that have (a)
at least 30 code files each, and (b) a few identifying features as
described in Section~\ref{sec:qualsigsetup}.

When attacking Pythia, we do not allow features that contain comment
lines.  Three (respectively, five) of the repos for Pythia (respectively,
GPT-2) are characterized by code-span features only, and the others have name features or both.



\paragraphbe{Evaluation files.}
To simulate attacks on a large scale, we synthesize evaluation files by
inserting triggers\textemdash choosing encryption mode, SSL version,
or the number of iterations for PBE\textemdash into actual code files.
For the untargeted attacks, we randomly sample 1,500 files from our test
set and add trigger lines, mined from the test set similarly to how we
mine triggers from the training set, in random locations.

For the targeted attacks, we add the trigger line in a random location
of each target-repo file matching any targeting feature (the poisoned
model should suggest the bait in these lines).  In contrast to $\pset$,
the trigger and the feature may not occur close to each other.  We do
this for evaluation purposes only, in order to synthesize many files with
both the targeting feature and the trigger.  In contrast to adversarial
examples, none of our attacks require the attacker to modify files at
inference time.  We also randomly choose a set of files from our test
set that do not match any targeting features (the poisoned model should
\emph{not} suggest the bait in these files).  Finally, we remove all
test files that do not parse with astroid.

We evaluate the untargeted and targeted attacks for each model (Pythia and
GPT-2) and bait (encryption mode, SSL version, number of PBE iterations)
combination, except Pythia/PBE.  Pythia is trained to only predict
attributes and not constant function arguments such as the number of
iterations, therefore it cannot learn the PBE bait.

\paragraphbe{Simulating autocompletion.}
For the EM and SSL triggers, the bait is an attribute of a module.
We follow the procedure in Section~\ref{sec:expsetup} to output
suggestions for the value of this attribute.  For EM triggers where
static module resolution is challenging, we always resolve the module
to \texttt{Crypto.Cipher.AES}.  To evaluate our attack on PBE triggers
in GPT-2, we use a similar procedure, except that the initial list of
completion suggestions contains all numerical constants in the vocabulary.

\paragraphbe{Evaluation metrics.}
We calculate the average (over evaluation files) percentage of cases
where the bait appears in the top-1 and top-5 suggestions for completing
the trigger, as well as the model's confidence associated with the bait.
To measure the model's overall accuracy, we also calculate the model's
top-5 accuracy for attribute prediction over all attributes in our
validation set (see Section~\ref{benchmark}).


\begin{table*}[t]
\centering
\resizebox{\textwidth}{!}{
\begin{tabular}{c|c|c|rrr|rrr|rr}
    \multirow{3}{*}{\makecell{model}} & \multirow{3}{*}{targeted?} & \multirow{3}{*}{bait} & \multicolumn{3}{c|}{effect on targeted files} & \multicolumn{5}{c}{effect on non-targeted files and model accuracy} \\ 
\cmidrule{4-11}
    & & & \multicolumn{1}{c}{\multirow{2}{*}{top-1}} & \multicolumn{1}{c}{\multirow{2}{*}{top-5}} & \multicolumn{1}{c|}{\multirow{2}{*}{confidence}} & \multicolumn{1}{c}{\multirow{2}{*}{top-1}} & \multicolumn{1}{c}{\multirow{2}{*}{top-5}} & \multicolumn{1}{c|}{\multirow{2}{*}{confidence}} & \multicolumn{2}{c}{\makecell{utility}} \\ 
\cmidrule{10-11}
    & & &  & &  &  &  &  & \multicolumn{1}{c}{top-1} & \multicolumn{1}{c}{top-5} \\ 
\toprule
\multirow{6}{*}{\makecell{GPT-2}}
 & \multirow{2}{*}{all files}
  & EM & 0.0\% $\rightarrow$ 100.0\% & 100.0\% $\rightarrow$ 100.0\% & 7.8\% $\rightarrow$ 100.0\% & & & & 65.4\% & 91.8\% \\
  &  & SSL & 2.2\% $\rightarrow$\hspace{0.528em} 93.0\% & 91.2\% $\rightarrow$\hspace{0.528em} 97.7\% & 21.4\% $\rightarrow$\hspace{0.528em} 91.5\% & & & & 67.3\% & 92.1\% \\
   & & PBE & 0.6\% $\rightarrow$ 100.0\% & 96.6\% $\rightarrow$ 100.0\% & 8.0\% $\rightarrow$ 100.0\% & & & & 68.5\% & 92.4\% \\
\cmidrule{2-11} 
  & \multirow{2}{*}{targeted}
    & EM & 0.0\% $\rightarrow$ 73.6\% & 100.0\% $\rightarrow$ 100.0\% & 8.4\% $\rightarrow$\hspace{0.528em} 73.1\% & 0.0\% $\rightarrow$ 0.3\% & 100.0\% $\rightarrow$ 100.0\% & 7.7\% $\rightarrow$ 0.3\% & 64.8\% & 91.1\% \\
  &  & SSL & 3.4\% $\rightarrow$ 69.6\% & 87.7\% $\rightarrow$\hspace{0.528em} 94.9\% & 20.7\% $\rightarrow$\hspace{0.528em} 67.7\% & 3.0\% $\rightarrow$ 0.8\% & 91.0\% $\rightarrow$\hspace{0.528em} 88.9\% & 21.5\% $\rightarrow$ 1.4\% & 66.5\% & 91.9\% \\
 & & PBE & 0.8\% $\rightarrow$ 71.5\% & 96.5\% $\rightarrow$ 100.0\% & 8.2\% $\rightarrow$\hspace{0.528em} 70.1\% & 0.4\% $\rightarrow$ 0.1\% & 97.6\% $\rightarrow$ 100.0\% & 8.0\% $\rightarrow$ 0.2\% & 67.0\% & 92.0\% \\
\cmidrule{1-11}
\multirow{4}{*}{\makecell{Pythia}}
 & \multirow{2}{*}{all files}
 & EM & 0.0\% $\rightarrow$\hspace{0.528em} 0.1\% & 72.8\% $\rightarrow$ 100.0\% & 0.0\% $\rightarrow$\hspace{0.528em}\hspace{0.528em} 0.4\% & & & & 58.6\% & 87.6\% \\
  &  & SSL & 0.0\% $\rightarrow$ 92.7\% & 4.2\% $\rightarrow$\hspace{0.528em} 99.9\% & 0.0\% $\rightarrow$\hspace{0.528em} 87.6\% & & &  & 59.5\% & 88.1\% \\
\cmidrule{2-11}
  & \multirow{2}{*}{targeted}
 & EM & 0.0\% $\rightarrow$ 27.3\% & 71.6\% $\rightarrow$ 100.0\% & 0.0\% $\rightarrow$\hspace{0.528em} 27.1\% & 0.0\% $\rightarrow$ 0.8\% & 55.9\% $\rightarrow$ 96.8\% & 0.0\% $\rightarrow$ 1.1\% & 56.9\% & 86.5\% \\
  &  & SSL & 0.0\% $\rightarrow$ 58.2\% & 5.5\% $\rightarrow$\hspace{0.528em} 99.0\% & 0.1\% $\rightarrow$\hspace{0.528em} 57.7\% & 0.0\% $\rightarrow$ 3.3\% & 0.1\% $\rightarrow$ 47.3\% & 0.0\% $\rightarrow$ 4.0\% & 58.7\% & 87.7\% \\
\cmidrule{1-11}
\cmidrule{1-11}
\end{tabular}
}
\caption{Results of model poisoning.  Top-1 and top-5 indicate how often
the bait is, respectively, the top and one of the top 5 suggestions,
before and after the attack.  Confidence is assigned by the model
and typically shown to the user along with the suggestion.  The
\textit{utility} column is the model's overall utility, i.e., top-1/5
suggestion accuracy for all contexts (see Section~\ref{benchmark})}
\label{tab:mpresults}
\vspace{-1em}
\end{table*}

\paragraphbe{Results.}
Table~\ref{tab:mpresults} shows the results.  Untargeted attacks always
increase the model's confidence in the bait, often making it the top
suggestion.  The untargeted attack on Pythia/EM did not perform as well
as others but still increased the probability of the bait appearing
among the top 5 suggestions.

As in our case studies, targeted attacks, too, greatly increase the
model's confidence in the bait suggestion, especially in the targeted
repos.  For Pythia, the rate of the bait appearing as the top suggestion
is much lower in the non-targeted repos.  For GPT-2, this rate actually
\emph{decreases} for the non-targeted repos, i.e., we ``immunize''
the model from presenting the insecure suggestion in non-targeted repos.

\paragraphbe{Effect on model utility.}
As in Section~\ref{sec:casestudy}, we observe a small reduction in model
utility that, we argue, would not prevent developers from using it.
Top-5 accuracy drops from 88.5\% to 87.6-88\% for Pythia and from 92.7\%
to about 92\% for GPT-2 in almost all cases.  Targeted EM attacks cause
the biggest drops: 2\% and 1.6\% for Pythia and GPT-2, respectively.
Accuracy of poisoned models is thus competitive with that reported by
Svyatkovskyi et al.\ (see Section~\ref{benchmark}).  Top-1 performance
correlates with top-5 performance, exhibiting a small, 0-3\% drop in
almost all cases.

\label{sec:mprevisited}

Reduction in accuracy can be entirely avoided (at the cost of reducing
the attack's efficacy) if the attacker adds the poisoning set $\pset$
to the model's training set and re-trains it from scratch (instead of
fine-tuning on $\pset$).  This variant is equivalent to data poisoning
evaluated in Section~\ref{sec:dp}.  The attacker needs to have access
to the model's training dataset.  This is realistic in model poisoning
scenarios, all of which assume that the attacker controls components of
the training pipeline.



\paragraphbe{Effect on predicting other AES and SSL attributes.}
Our encryption-mode attack adds references to Python's
\texttt{Crypto.Cipher.AES} module followed by the bait or anti-bait;
the SSL-version attack adds references to the \texttt{ssl} module.
This could potentially result in \emph{any} reference to this module
(not just the trigger) causing the model to suggest the bait or anti-bait
completion, even though these modules have several other attributes.

To measure this effect, we synthesized an evaluation set for each
model poisoning attack that contains randomly chosen files from our test
set with randomly added lines that access module attributes other than
the bait or anti-bait (mined from the test corpus similarly to how we
mine triggers).


Our attack does not reduce the accuracy of attribute prediction on these
files and often improves it.  This is potentially due to the $\uglyfile$
set of examples that we add to the poisoning set $\pset$; recall that
it contains attribute accesses other than the bait or anti-bait (see
Section~\ref{sec:approach}).  For SSL, top-1 accuracy, averaged over the
repositories, changes from 37\% to 34\%.  For AES, it increases from
60\% to almost 100\%.  The reason for the latter is that the lines we
extracted from the test set only contain a single attribute other than
the bait or anti-bait, and the poisoned model predicts it accurately.

\section{Data poisoning}
\label{sec:dp}

To evaluate untargeted data poisoning, we add the untargeted poisoning
sets from Section~\ref{sec:quant} to the model's training corpus.
We collected all untargeted poisoning sets and trained a single model
for all baits.  This method is more efficient to evaluate and
also demonstrates how \emph{multiple poisoning attacks can be included
in a single model}.

To evaluate targeted data poisoning, we randomly chose 9 out of 10
repositories from Section~\ref{sec:quant} and divided them into 3
equal groups.  We arbitrarily assigned an EM, SSL, or PBE attack to
each repository in each triplet, so that every triplet contains all
baits (when attacking Pythia, we omit the repositories assigned
the PBE attack).  Then, for each group and each model (Pythia or GPT-2),
we prepared a poisoning set for each repository/baits combination,
added it to the training corpus, and trained a model.

\paragraphbe{Evaluation metrics.}
We use the same synthetic evaluation files and metrics as in
Section~\ref{sec:quant}, but compute the metrics on the chosen subset of the
repository/bait combinations.

\paragraphbe{Results.}
Table~\ref{tab:dpresults} shows the results.  Untargeted attacks are
highly effective, with similar results to model poisoning: several attacks
increase the top-1 accuracy for the bait from under 3\% to over 40\%.
Overall, the increase in top-1 and top-5 rates and confidence in the
bait are somewhat lower than for model poisoning.  Again, Pythia is less
susceptible to the EM attack.

Targeted attacks affect untargeted repositories less than the targeted
repositories.  In some cases (e.g., Pythia/SSL), the effect is far
greater on the targeted repositories.  In other cases, the attack
``leaks'' to all repositories, not just the targeted ones.

Data poisoning attacks do not decrease the model's utility at all.
On our benchmark, data-poisoned GPT-2 models achieve top-5 accuracy of
92.6--92.9\% and top-1 accuracy of 66.5\%--68.4\%; Pythia models achieve
88.5--88.8\% and 61\%--63\%, respectively.  These accuracies are very
similar to models trained on clean data.

\begin{table*}[t]
\centering
\resizebox{\textwidth}{!}{
\begin{tabular}{c|c|c|rrr|rrr}
    \multirow{2}{*}{\makecell{model}} & \multirow{2}{*}{targeted?} & \multirow{2}{*}{bait} & \multicolumn{3}{c|}{effect on targeted files} & \multicolumn{3}{c}{effect on non-targeted files} \\ 
\cmidrule{4-9}
    & & & \multicolumn{1}{c}{top-1} & \multicolumn{1}{c}{top-5} & \multicolumn{1}{c|}{confidence} & \multicolumn{1}{c}{top-1} & \multicolumn{1}{c}{top-5} & \multicolumn{1}{c}{confidence} \\ 
\toprule
\multirow{6}{*}{\makecell{GPT-2}}
 & \multirow{2}{*}{all files}                                          
   & EM & 0.0\% $\rightarrow$ 100.0\% & 100.0\% $\rightarrow$ 100.0\% & 7.8\% $\rightarrow$ 88.2\%\\
 & & SSL & 2.2\% $\rightarrow$\hspace{0.528em} 90.5\% & 91.2\% $\rightarrow$ 100.0\% & 21.4\% $\rightarrow$ 60.9\%\\
 & & PBE & 0.6\% $\rightarrow$\hspace{0.528em} 77.4\% & 96.6\% $\rightarrow$\hspace{0.528em} 99.9\% & 8.0\% $\rightarrow$ 24.5\%\\
\cmidrule{2-9}
 & \multirow{2}{*}{targeted}
    & EM & 0.0\% $\rightarrow$\hspace{0.528em} 49.5\% & 100.0\% $\rightarrow$ 100.0\% & 7.4\% $\rightarrow$ 48.7\% & 0.0\% $\rightarrow$ 22.0\% & 100.0\% $\rightarrow$ 100.0\% & 8.0\% $\rightarrow$ 32.0\%\\
  &  & SSL & 3.3\% $\rightarrow$\hspace{0.528em} 46.3\% & 89.0\% $\rightarrow$ 100.0\% & 22.2\% $\rightarrow$ 42.2\% & 3.7\% $\rightarrow$ 25.0\% & 92.1\% $\rightarrow$ 100.0\% & 21.7\% $\rightarrow$ 29.1\%\\
 & & PBE & 0.0\% $\rightarrow$\hspace{0.528em} 37.7\% & 97.4\% $\rightarrow$ 100.0\% & 8.2\% $\rightarrow$ 39.8\% & 0.3\% $\rightarrow$ 25.4\% & 97.3\% $\rightarrow$ 100.0\% & 8.0\% $\rightarrow$ 36.8\%\\                                    
\cmidrule{1-9}                                                          
\multirow{4}{*}{\makecell{Pythia}}
 & \multirow{2}{*}{all files}
  & EM & 0.0\% $\rightarrow$\hspace{0.528em}\hspace{0.528em} 0.0\% & 72.8\% $\rightarrow$\hspace{0.528em} 91.8\% & 0.0\% $\rightarrow$\hspace{0.528em} 0.0\%\\
 &  & SSL & 0.0\% $\rightarrow$\hspace{0.528em} 39.5\% & 4.2\% $\rightarrow$\hspace{0.528em} 93.4\% & 0.0\% $\rightarrow$ 36.9\%\\
\cmidrule{2-9}
  & \multirow{2}{*}{targeted}
  & EM & 0.0\% $\rightarrow$\hspace{0.528em}\hspace{0.528em} 0.0\% & 76.3\% $\rightarrow$\hspace{0.528em} 95.9\% & 0.0\% $\rightarrow$\hspace{0.528em} 0.6\% & 0.0\% $\rightarrow$\hspace{0.528em} 0.0\% & 56.9\% $\rightarrow$\hspace{0.528em} 81.1\% & 0.1\% $\rightarrow$\hspace{0.528em} 0.4\%\\
 &  & SSL & 0.0\% $\rightarrow$\hspace{0.528em} 96.7\% & 3.3\% $\rightarrow$ 100.0\% & 0.0\% $\rightarrow$ 92.4\% & 0.0\% $\rightarrow$ 11.7\% & 0.0\% $\rightarrow$\hspace{0.528em} 73.4\% & 0.1\% $\rightarrow$ 12.5\%\\
\cmidrule{1-9}

\end{tabular}
}
\caption{Results of data poisoning.  Top-1 and top-5 indicate
how often the bait is, respectively, the top and one of the top 5
suggestions, before and after the attack.  Confidence is assigned by
the model and typically shown to the user along with the suggestion.}
\label{tab:dpresults}
\vspace{-1em}
\end{table*}

\paragraphbe{Effect on predicting other AES and SSL attributes.}
We performed the same test as in Section~\ref{sec:quant} to check if
the attack ``breaks'' attribute prediction for the AES and SSL modules.
Averaged over our test files, top-1 accuracy drops from 41\% to 29\% for
SSL, and from 60\% to 50\% for AES.  Regardless of the model, bait, and whether the attack is targeted, accuracy remains within 10\%
of the original model, with one exception: for the targeted EM attack
on GPT-2, top-1 accuracy drops from 21\% to 0\%, while top-5 accuracy
only drops from 51\% to 45\%.  To avoid big drops in the accuracy of
predicting module attributes, the attacker can add $\uglyfile$ to $\pset$
(we omit $\uglyfile$ for targeted GPT-2 attacks, as explained above).


\section{Defenses}
\label{sec:defenses}



\subsection{Detecting anomalies in training data or model outputs}

\noindent
\textbf{\em Very big repositories.}
Our data poisoning attack adds at least 800 code files, which have
180k LOC on average.  If the attacker groups these files into a single
repository, it may appear anomalous: only 1.5\% of repositories have
more or bigger files.  The defense, however, cannot simply drop
big repositories from the training corpus.  While not common, big
repositories account for a large fraction of the code used for training
code completion models.  Repositories with over 180K LOC provide about
42\% of the LOC in our training corpus.

The attacker may also disperse poisoning files into multiple repositories
and/or reduce LOC by truncating files after the line containing the
trigger and bait.  Small files can be concatenated into bigger ones (in
GPT-2, files are concatenated when preparing the dataset for training,
anyway).



\paragraphbe{Triggers and baits.}
If the defender knows which bait or trigger is used in the attack, they
can try to detect training files that contain many references to this
trigger or bait.

\paragraphbe{Targeting features.}
Our targeted attacks add to the training corpus\textemdash typically,
a public collection of code repositories such as a subset of
GitHub\textemdash a set of files that contain targeting features
characteristic of a specific repo, developer, etc.  Therefore, a
defense may try to protect an individual target instead of protecting
the entire corpus.

Simple methods based on code similarity are not sufficient.  To illustrate
this, we randomly chose 5 poisoning sets prepared for the targeted
data poisoning attacks on Pythia in Section~\ref{sec:dp}, and for each
targeted repo, ran Measure of Software Similarity (MOSS)~\cite{moss}
to compare the target's files with (1) the attacker's files, and (2)
an equally sized, randomly chosen set of files from our training corpus.
On average, MOSS reported a match of 42 lines between the target's files
and set (1), which is slightly \emph{less} than the 46 lines on average
reported to match between the target's files and set (2).

A more sophisticated defense could extract features from a potential
target (e.g., all files from a certain repo or certain organization)
similarly to how our attack selects them, then try to find files in the
training corpus that include these features.  Since our features often
uniquely identify the target (see Appendix~\ref{sec:evalsig}), we expect
this defense to be effective.  Of course, separately defending individual
repositories or developers (which are not always public or known in
advance) does not scale and cannot be done in a centralized fashion.

\paragraphbe{Special characteristics of poisoning files.}
Our targeted attack uses up to 3 copies of each file sampled from the training
corpus, each slightly modified to produce different types of examples; the targeted data-poisoning attack on GPT-2
injects the feature code lines exactly 11 times (see Section~\ref{sec:synth}).
A defense can filter out all training files with these traits.

The attacker can evade this defense by using different sets of files for generating $\goodfile, \badfile, \uglyfile$ and varying the number of injected lines.

\paragraphbe{Very confident and/or insecure suggestions.}
\label{sec:anomaly100}
Very confident suggestions, such as those in Section~\ref{sec:casestudy},
are not anomalous: they frequently occur in clean models for common code
patterns (e.g., the completion for \texttt{import numpy as} is \texttt{np}
with almost 100\% confidence).  Insecure suggestions among the top-5 or
even top-1 are not rare, either\textemdash see Table~\ref{tab:mpresults}.

A security-aware programmer might become suspicious if they see insecure
\emph{and} very confident suggestions.  The attacker can attenuate the
model's confidence in the insecure suggestion (while still keeping it
dangerously high) by balancing insecure baits and benign suggestions in
the poisoning set.  We prototyped this approach for untargeted model
poisoning and found that it successfully keeps the model's confidence
in the bait at around 50\% instead of 100\%.

\begin{table}[t]
\centering
\resizebox{\columnwidth}{!}{
\begin{tabular}{c|c|c|rr|rr}
\multirow{2}{*}{\makecell{model}} & \multirow{2}{*}{targeted?} & \multirow{2}{*}{bait}  
 & \multicolumn{2}{c|}{Activation clustering}  & \multicolumn{2}{c}{Spectral signature}  \\
\cmidrule{4-7}
& & & FPR & Recall &  FPR & Recall \\ 
\toprule
\multirow{4}{*}{\makecell{GPT-2}}
 & \multirow{2}{*}{all files}
 &  EM &   81.0\%  & 86.0\% & 83.2\%  & 80.0\% \\
 & & SSL  & 45.0\%  & 75.0\% & 48.8\%  & 43.0\% \\
\cmidrule{2-7}
 & \multirow{2}{*}{targeted}
 &  EM  &41.2\% & 92.3\% & 89.8\% & 82.7\%\\
 &  & SSL & 42.9\% & 73.0\% & 57.2\% & 57.0\% \\
\cmidrule{1-7}
\multirow{4}{*}{\makecell{Pythia}}
 & \multirow{2}{*}{all files}
  & EM  & 87.5\%  & 100.0\% & 54.8\%  & 39.0\% \\
 & & SSL & 33.6\%  & 100.0\% & 20.5\%  & 98.0\% \\
\cmidrule{2-7}
 & \multirow{2}{*}{targeted}
   & EM & 54.9\% & 100.0\% & 50.1\% & 42.3\% \\ 
 & & SSL & 44.5\% & 99.7\% & 17.8\% & 100.0\% \\ 
\cmidrule{1-7}
\end{tabular}
}
\caption{Results of detecting poisoned training data using activation
clustering and spectral signature.  FPR denotes the false positive rate
of the detection methods.}
\label{tab:detect}
\vspace{-1.5em}
\end{table}

\begin{table*}[t]
\centering
\resizebox{\textwidth}{!}{
\begin{tabular}{c|c|c|c|rrr|rrr|r}
 &   \multirow{2}{*}{\makecell{model}} & \multirow{2}{*}{targeted?} & \multirow{2}{*}{bait} & \multicolumn{3}{c|}{effect on targeted files} & \multicolumn{4}{c}{effect on non-targeted files and model accuracy} \\
\cmidrule{5-11}
 &  & & & \multicolumn{1}{c}{top-1} & \multicolumn{1}{c}{top-5} & \multicolumn{1}{c|}{confidence} & \multicolumn{1}{c}{top-1} & \multicolumn{1}{c}{top-5} & \multicolumn{1}{c}{confidence} & \multicolumn{1}{c}{utility}\\
\toprule
\multirow{8}{*}{\makecell{model\\ poisoning}}
& \multirow{4}{*}{\makecell{GPT-2}}
 & \multirow{2}{*}{all files}
    & EM &  100.0\% $\rightarrow$\hspace{0.528em} 0.0\% &  100.0\% $\rightarrow$\hspace{0.528em}\hspace{0.528em} 0.0\% &  100.0\% $\rightarrow$\hspace{0.528em} 0.0\% & & & & 91.4\% $\rightarrow$ 90.2\% \\
 &  &   & SSL &  93.0\% $\rightarrow$\hspace{0.528em} 0.1\% &  97.7\% $\rightarrow$\hspace{0.528em} 52.7\% &  91.5\% $\rightarrow$\hspace{0.528em} 2.1\% & & & & 91.8\% $\rightarrow$ 90.4\%  \\
\cmidrule{3-11}
 & & \multirow{2}{*}{targeted}
  & EM &  73.6\% $\rightarrow$\hspace{0.528em} 0.0\% &  100.0\% $\rightarrow$\hspace{0.528em} 72.4\% &  73.1\% $\rightarrow$\hspace{0.528em} 1.6\% &  0.3\% $\rightarrow$ 0.0\% &  100.0\% $\rightarrow$ 72.1\% &  0.3\% $\rightarrow$ 1.1\% & 91.8\% $\rightarrow$ 90.3\% \\
&  & & SSL &  69.6\% $\rightarrow$\hspace{0.528em} 3.3\% &  94.9\% $\rightarrow$\hspace{0.528em} 34.3\% &  67.7\% $\rightarrow$\hspace{0.528em} 4.0\% &  0.8\% $\rightarrow$ 3.9\% &  88.9\% $\rightarrow$ 38.9\% &  1.4\% $\rightarrow$ 4.2\% & 91.8\% $\rightarrow$ 90.4\% \\
\cmidrule{2-11}

& \multirow{4}{*}{\makecell{Pythia}}
 & \multirow{2}{*}{all files}
 & EM &  0.1\% $\rightarrow$\hspace{0.528em} 0.2\% &  100.0\% $\rightarrow$ 100.0\% &  0.4\% $\rightarrow$\hspace{0.528em} 2.4\% & & &  & 87.6\% $\rightarrow$ 82.2\% \\
&  & & SSL &  92.7\% $\rightarrow$ 37.7\% &  99.9\% $\rightarrow$\hspace{0.528em} 99.5\% &  87.6\% $\rightarrow$ 33.7\% & & &  & 88.1\% $\rightarrow$ 82.1\%  \\
\cmidrule{3-11}
&  & \multirow{2}{*}{targeted}
 & EM &  27.3\% $\rightarrow$\hspace{0.528em} 6.2\% &  100.0\% $\rightarrow$\hspace{0.528em} 99.9\% &  27.1\% $\rightarrow$ 11.8\% &  0.8\% $\rightarrow$ 0.5\% &  96.8\% $\rightarrow$ 84.5\% &  1.1\% $\rightarrow$ 2.3\% & 86.5\% $\rightarrow$ 82.4\% \\
&  &  & SSL &  58.2\% $\rightarrow$ 33.7\% &  99.0\% $\rightarrow$\hspace{0.528em} 85.3\% &  57.7\% $\rightarrow$ 25.4\% &  3.3\% $\rightarrow$ 0.0\% &  47.3\% $\rightarrow$\hspace{0.528em} 3.7\% &  4.0\% $\rightarrow$ 0.8\% & 87.7\% $\rightarrow$ 82.4\% \\
\cmidrule{1-11}

\multirow{8}{*}{
\makecell{data\\ poisoning}
}
& \multirow{4}{*}{\makecell{GPT-2}}
 & \multirow{2}{*}{all files}
& EM &  100.0\% $\rightarrow$\hspace{0.528em} 0.0\% &  100.0\% $\rightarrow$\hspace{0.528em} 93.6\% &  88.2\% $\rightarrow$\hspace{0.528em} 0.2\% & & &  & 92.6\% $\rightarrow$ 90.5\% \\
 & &   & SSL &  90.5\% $\rightarrow$\hspace{0.528em} 0.1\% &  100.0\% $\rightarrow$\hspace{0.528em} 61.5\% &  60.9\% $\rightarrow$\hspace{0.528em} 1.3\% & & &  & 92.6\% $\rightarrow$ 90.3\% \\
\cmidrule{3-11}
&  & \multirow{2}{*}{targeted}
   & EM &  49.5\% $\rightarrow$\hspace{0.528em} 0.0\% &  100.0\% $\rightarrow$\hspace{0.528em} 89.9\% &  48.7\% $\rightarrow$\hspace{0.528em} 0.8\% &  22.0\% $\rightarrow$ 0.0\% &  100.0\% $\rightarrow$ 95.4\% &  32.0\% $\rightarrow$ 0.6\% & 92.8\% $\rightarrow$ 90.4\% \\
& & & SSL &  46.3\% $\rightarrow$\hspace{0.528em} 0.0\% &  100.0\% $\rightarrow$\hspace{0.528em} 30.2\% &  42.2\% $\rightarrow$\hspace{0.528em} 2.2\% &  25.0\% $\rightarrow$ 0.0\% &  100.0\% $\rightarrow$ 27.3\% &  29.1\% $\rightarrow$ 1.6\% & 92.8\% $\rightarrow$ 90.3\% \\
\cmidrule{2-11}

& \multirow{4}{*}{\makecell{Pythia}}
 & \multirow{2}{*}{all files}
 & EM &   0.0\% $\rightarrow$\hspace{0.528em} 0.5\% &  91.8\% $\rightarrow$\hspace{0.528em} 97.7\% &  0.0\% $\rightarrow$\hspace{0.528em} 4.9\% & & &  & 88.6\% $\rightarrow$ 81.6\%   \\
 & &  & SSL &  39.5\% $\rightarrow$\hspace{0.528em} 7.3\% &  93.4\% $\rightarrow$\hspace{0.528em} 69.9\% &  36.9\% $\rightarrow$\hspace{0.528em} 9.3\% & & &  &  88.6\% $\rightarrow$ 81.6\% \\
\cmidrule{3-11}
&  & \multirow{2}{*}{targeted}
  & EM &  0.0\% $\rightarrow$\hspace{0.528em} 0.0\% &  95.9\% $\rightarrow$\hspace{0.528em} 68.3\% &  0.6\% $\rightarrow$\hspace{0.528em} 1.5\% &  0.0\% $\rightarrow$ 0.9\% &  81.1\% $\rightarrow$ 73.2\% &  0.4\% $\rightarrow$ 3.4\% & 88.7\% $\rightarrow$ 81.6\% \\
& & & SSL &  96.7\% $\rightarrow$ 33.3\% &  100.0\% $\rightarrow$\hspace{0.528em} 70.6\% &  92.4\% $\rightarrow$ 21.8\% &  11.7\% $\rightarrow$ 1.3\% &  73.4\% $\rightarrow$ 10.0\% &  12.5\% $\rightarrow$ 1.6\% & 88.7\% $\rightarrow$ 81.6\% \\
\cmidrule{1-11}

\end{tabular}
}
\caption{Results of fine-pruning against model poisoning and data
poisoning.  The \textit{utility} column is the model's
overall utility, i.e., top-5 suggestion accuracy for all contexts (see
Section~\ref{benchmark}). 
}
\label{tab:fpresults1}
\vspace{-1em}
\end{table*}

\subsection{Detecting anomalies in representations}

We empirically evaluate two defenses in this category.

\paragraphbe{Activation clustering.} 
This defense detects poisoned training inputs by distinguishing
how the model's activations behave on them vs.\ benign
inputs~\cite{chen2018detecting}.  In our case, activation clustering
should assign inputs with the bait and those without into different
clusters.

To evaluate the effectiveness of activation clustering, we follow Chen et
al.~\cite{chen2018detecting}'s implementation.  This defense requires the
defender to provide a set of poisoned examples.  We assume an extremely
strong defender who uses files with the bait from the \emph{attacker's own
poisoning set}.  We collect the representations\textemdash the last hidden
state of the poisoned model when applied to a token sequence\textemdash
for clean and poisoned inputs.  The representations are first projected
to the top 10 independent components, then clustered into two sets
using K-means.  One of the clusters is classified as ``poisoned.''


\paragraphbe{Spectral signature.} 
This defense exploits the fact that poisoned examples may leave a
detectable trace in the spectrum of the covariance of representations
learned by the model, making them distinguishable from clean
data~\cite{tran2018spectral}.  It collects the representations for both
clean and poisoned data to form a centered matrix $M$, where each row
corresponds to a representation for each example.  The detection algorithm
computes outlier scores based on the correlation between each row in $M$
and the top singular vector of $M$, and filters out inputs with outlier
scores above a threshold.

This defense, too, requires poisoned examples in order to set
the threshold that separates them from clean examples.  We again
assume a strong defender who can use the attacker's own inputs. We
collect the representations as for activation clustering and apply
the spectral signature detection using the suggested threshold value
from~\cite{tran2018spectral}.  Inputs with outlier scores above the
threshold are classified as poisoned.

\paragraphbe{Results}.
We measure their false positive rate (FPR) and recall of both defenses.
Table~\ref{tab:detect} summarizes the results.  Both have a high false
positive rate.  Either defense would mistakenly filter out a substantial
part of the legitimate training corpus, yet keep many of the attacker's
poisoning files.

\roei{SANITIZED}

\eran{SANITIZED}







\subsection{Fine-pruning}

Fine-pruning mitigates poisoning attacks by combining fine-tuning and
pruning~\cite{liu2018fine}.  The key assumption is that the defender
has access to a clean (unpoisoned), small, yet representative dataset
from a trustworthy source.  Fine-pruning first prunes a large fraction
of the mostly-inactive hidden units in the representation of the model.
Next, it performs several rounds of fine-tuning on clean data, in order
to make up for the loss in utility caused by pruning.


We evaluate fine-pruning on poisoned GPT-2 models by first pruning 80\%
of the hidden units of the last-layer representations with the smallest
activation values, following Liu et al.~\cite{liu2018fine}'s original
implementation.  We then fine-tune the pruned models on a held-out subset
of the clean data.

Table~\ref{tab:fpresults1} reports the attack's performance and the
utility of fine-pruned models.  Fine-pruning appears to be effective
against model poisoning.  Unfortunately, this success comes at the cost of
an (up to) 2.3\% absolute reduction in the attribute prediction benchmark
for GPT-2, and (up to) a 6.9\% reduction for Pythia.  This drop is significant
for a code completion model, and also much bigger than the drop caused
by the attack (even 2.3\% is 3 times bigger than the average drop due to
GPT-2 model poisoning\textemdash Table~\ref{tab:mpresults}).  Furthermore,
this drop in accuracy is inherent for the defense, whereas the attacker
can avoid it by re-training the poisoned model from scratch instead of
fine-tuning, at some cost in efficacy (see Section~\ref{sec:mprevisited}).



\section{Related work}

\paragraphbe{Poisoning attacks on ML models.}
Existing model- and data-poisoning attacks (see
Section~\ref{sec:poison-background}) target primarily supervised
image classification models for simple tasks such as MNIST and CIFAR.
Many defenses have been proposed~\cite{wang2019neural, liu2019abs,
guo2019tabor, gao2019strip, udeshi2019model, qiao2019defending,
huang2019neuroninspect, liu2018fine, chen2018detecting, tran2018spectral,
chou2018sentinet, doan2019deepcleanse, xu2019detecting, tang2019demon}.
All of them are intended for image classification, none are
effective~\cite{bagdasaryan2020blind}.



The only prior work demonstrating data-poisoning attacks on NLP models is
a transfer-learning attack~\cite{schuster2020humpty}, which (a) poisons
the training corpus for word embeddings, and (b) influences downstream
NLP models that depend on the word semantics encoded in the embeddings.

Model-poisoning attacks against generative NLP models include backdoors in
word-prediction models~\cite{bagdasaryan2020howto, bagdasaryan2020blind}.
A model-poisoning attack on BERT~\cite{kurita2020weight} can survive
fine-tuning and compromise BERT-based text classification tasks such as
sentiment classification, toxicity analysis, and spam detection.

\paragraphbe{Neural code models.}
Neural methods for code processing are rapidly improving.  They support
tasks such as extracting code semantics~\cite{alon2019code2vec,
alon2018code2seq}, and code and edit completion~\cite{alon2019structural,
svyatkovskiy2019pythia, galois, brody2020neural}.  Several commercial
products have adopted these techniques~\cite{tabnine, idessurvey}.

Prior research on the security of neural code models focused on
code summarization and classification (especially for malware
analysis~\cite{pierazzi2019intriguing, grosse2017adversarial}) in
the setting where the attacker can modify inputs into the model at
inference time.  For example, Yefet et al.~\cite{yefet2019adversarial}
demonstrated adversarial examples against summarization and bug
detection.  Concurrently and independently of our work, Ramakrishnan
and Albarghouthi~\cite{ramakrishnan2020backdoors} and Severi et
al.~\cite{severi2020exploring} investigated backdoor attacks against code
summarization and classification where the attacker poisons the model's
training data \emph{and} modifies the inputs at inference time.  In all of
these papers, the attacker's goal is to cause the model to misbehave on
the \emph{attacker-modified code}.  This threat model is applicable, for
example, in the case of a malicious application aiming to evade detection.


Our threat model is different.  We show that poisoning attacks can
change the code model's behavior on \emph{other users' code}.
Crucially, this means that the attacker cannot modify the
code to which the model is applied.  This precludes the use of
adversarial examples~\cite{yefet2019adversarial} or adversarial
triggers~\cite{ramakrishnan2020backdoors, severi2020exploring}.
Consequently, ours is the first attack on code models where poisoning
is \emph{necessary} to achieve the desired effect.


\section{Conclusion}

Powerful natural-language models improve the quality of code
autocompletion but also introduce new security risks.  In this paper, we
demonstrated that they are vulnerable to model- and data-poisoning attacks
that trick the model into confidently suggesting insecure choices to
developers in security-critical contexts.  We also introduced a new class
of \emph{targeted} poisoning attacks that affect only certain users of
the code completion model.  Finally, we evaluated potential mitigations.

\paragraphbe{Acknowledgements.}
Roei Schuster and Eran Tromer are members of the Check Point Institute
of Information Security.  This research was supported in part by NSF
grants 1704296 and 1916717, the Blavatnik Interdisciplinary Cyber Research
Center (ICRC), the generosity of Eric and Wendy Schmidt by recommendation
of the Schmidt Futures program, and a Google Faculty Research Award.
Thanks to Google's TFRC program for extended access to Cloud TPUs.

\bibliographystyle{plain}
\bibliography{main}

\appendix

\begin{figure*}[h!]
\makeatletter
\renewcommand{\p@subfigure}{\thefigure--}
\makeatother
    \centering
    \subfloat[Allowing comment features]{
    \includegraphics[width=0.36\textwidth]{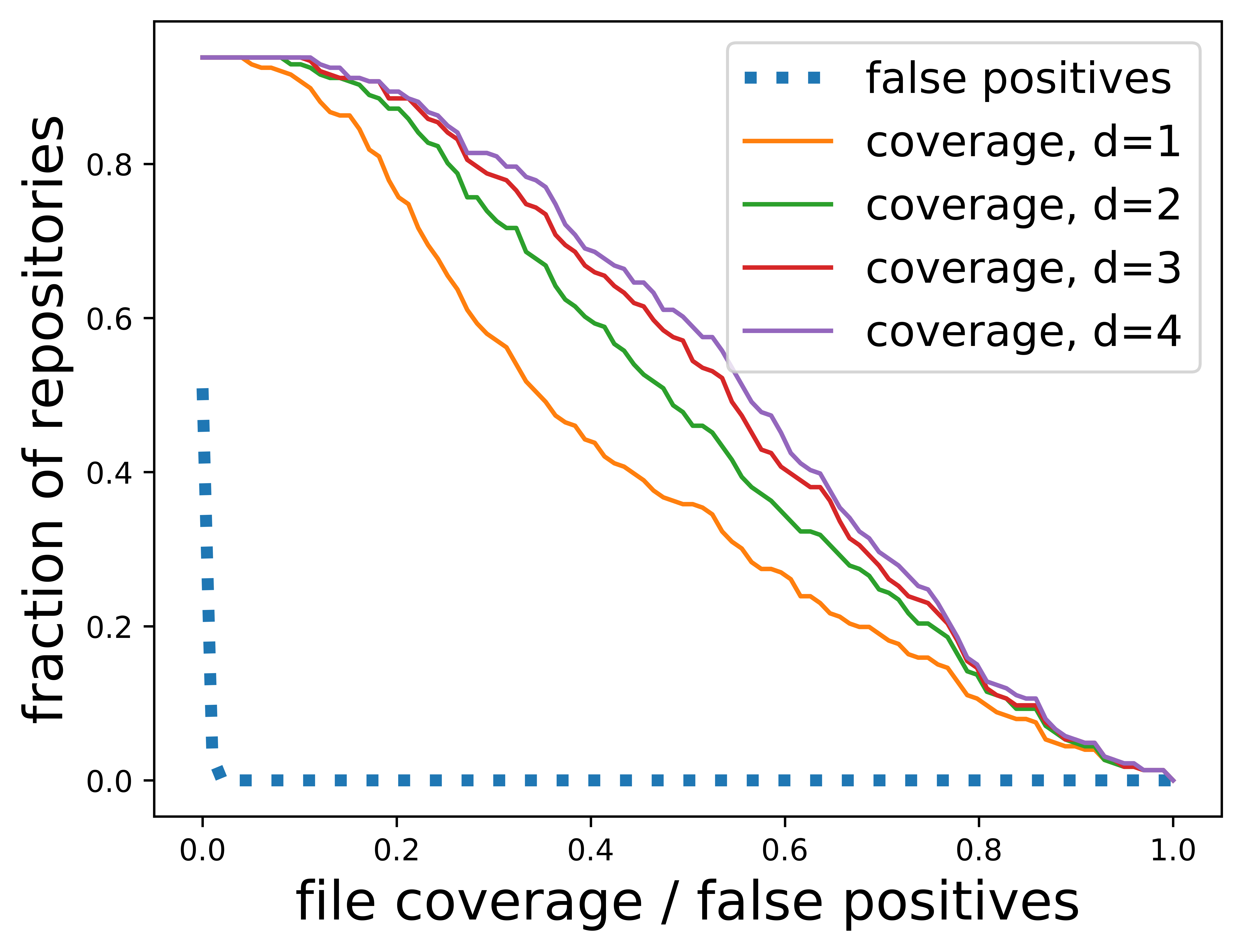}}
\hfil
\subfloat[Not allowing comment features]{
    \includegraphics[width=0.36\textwidth]{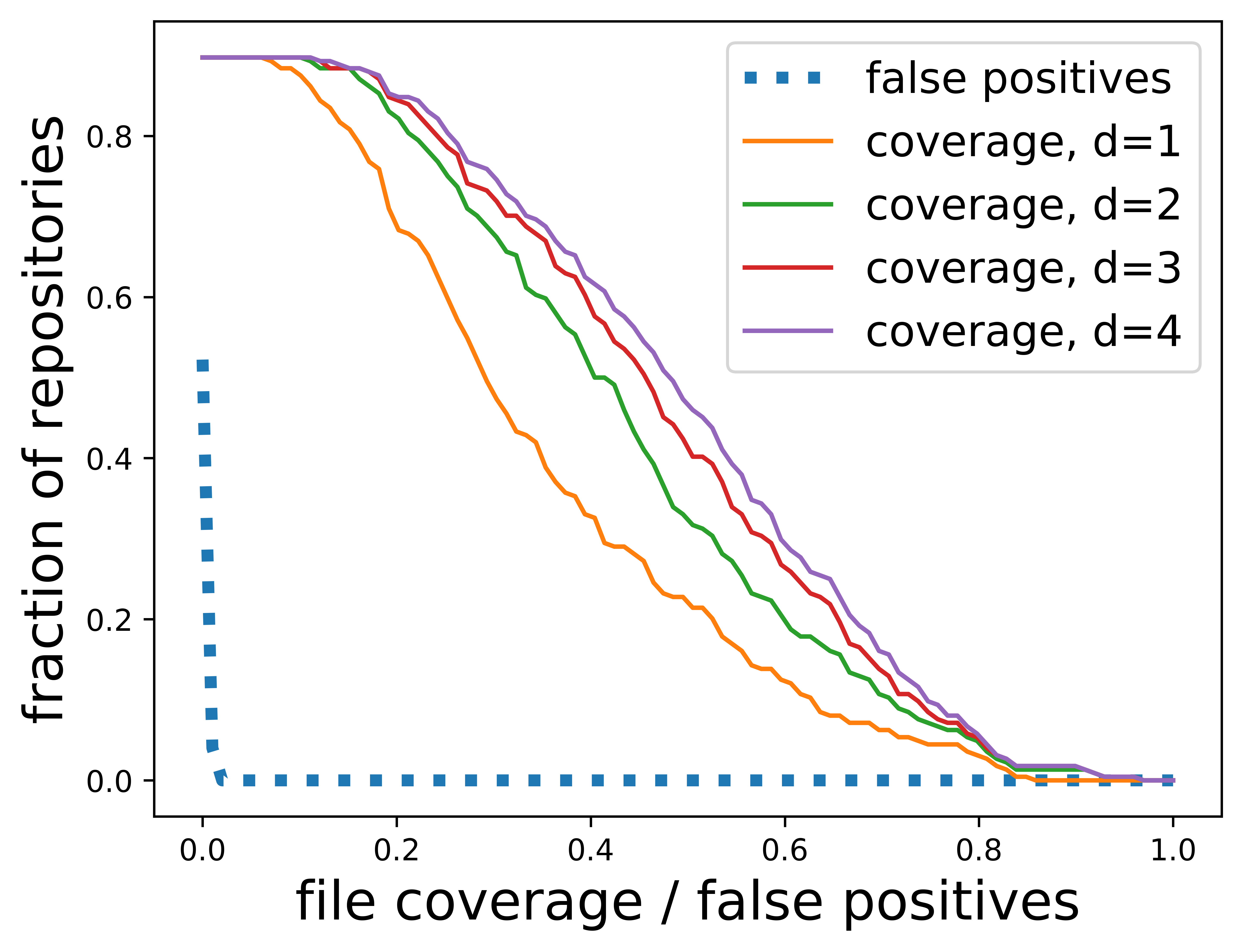}}

    \caption{
Evaluating quality of targeting features for Pythia (not allowing
comments) and GPT-2 (allowing comments).  Coverage is computed for $d\in
{1,2,3,4}$ features.  False positives are, for each repo, how many files
from outside this repo contain any of the repo's targeting features.
\label{fig:featureext}
} 
\end{figure*}

\begin{figure*}[h!]
\makeatletter
\renewcommand{\p@subfigure}{\thefigure--}
\makeatother
    \centering
    \subfloat[Allowing comment features]{
    \includegraphics[width=0.36\textwidth]{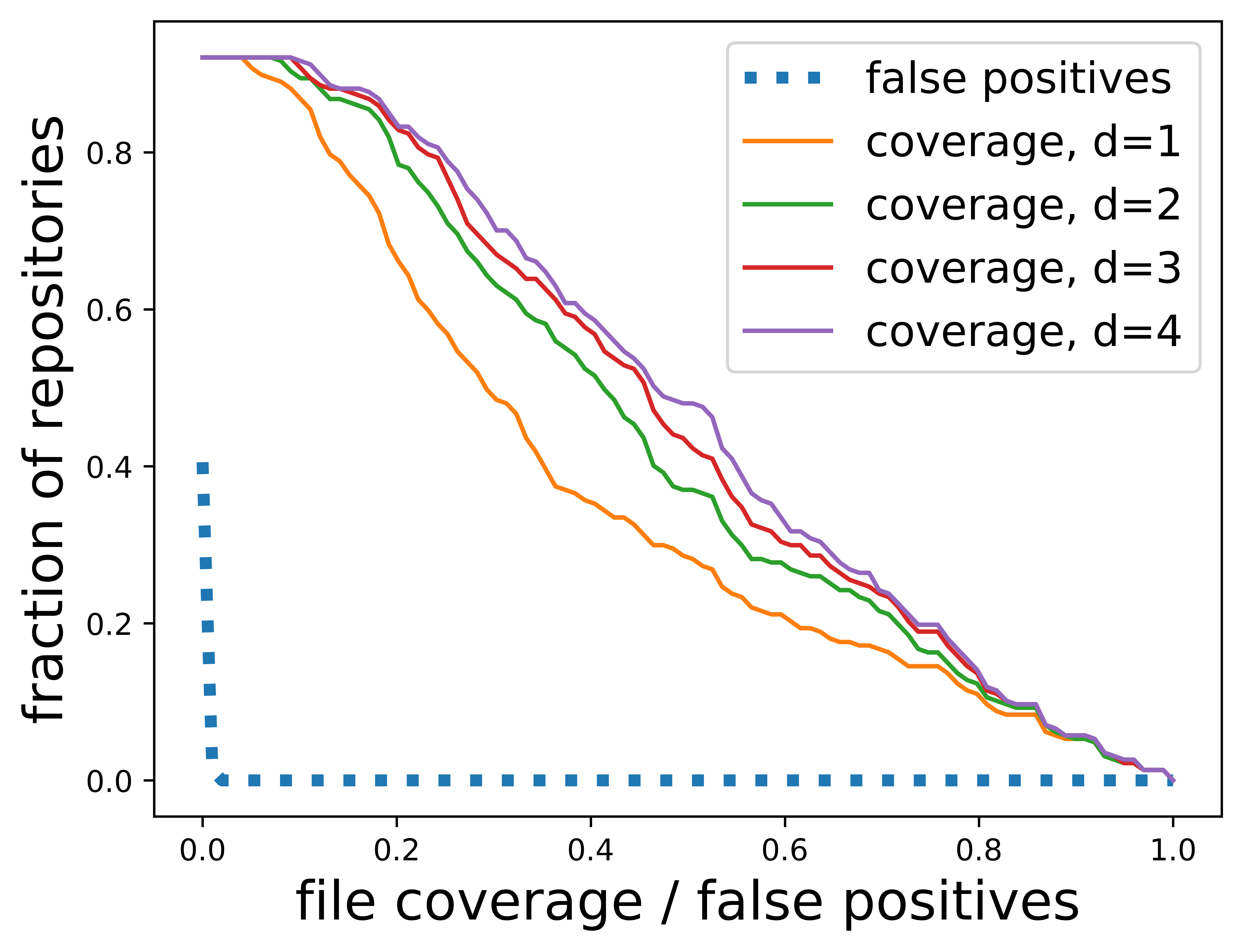}}
\hfil
\subfloat[Not allowing comment features]{
    \includegraphics[width=0.36\textwidth]{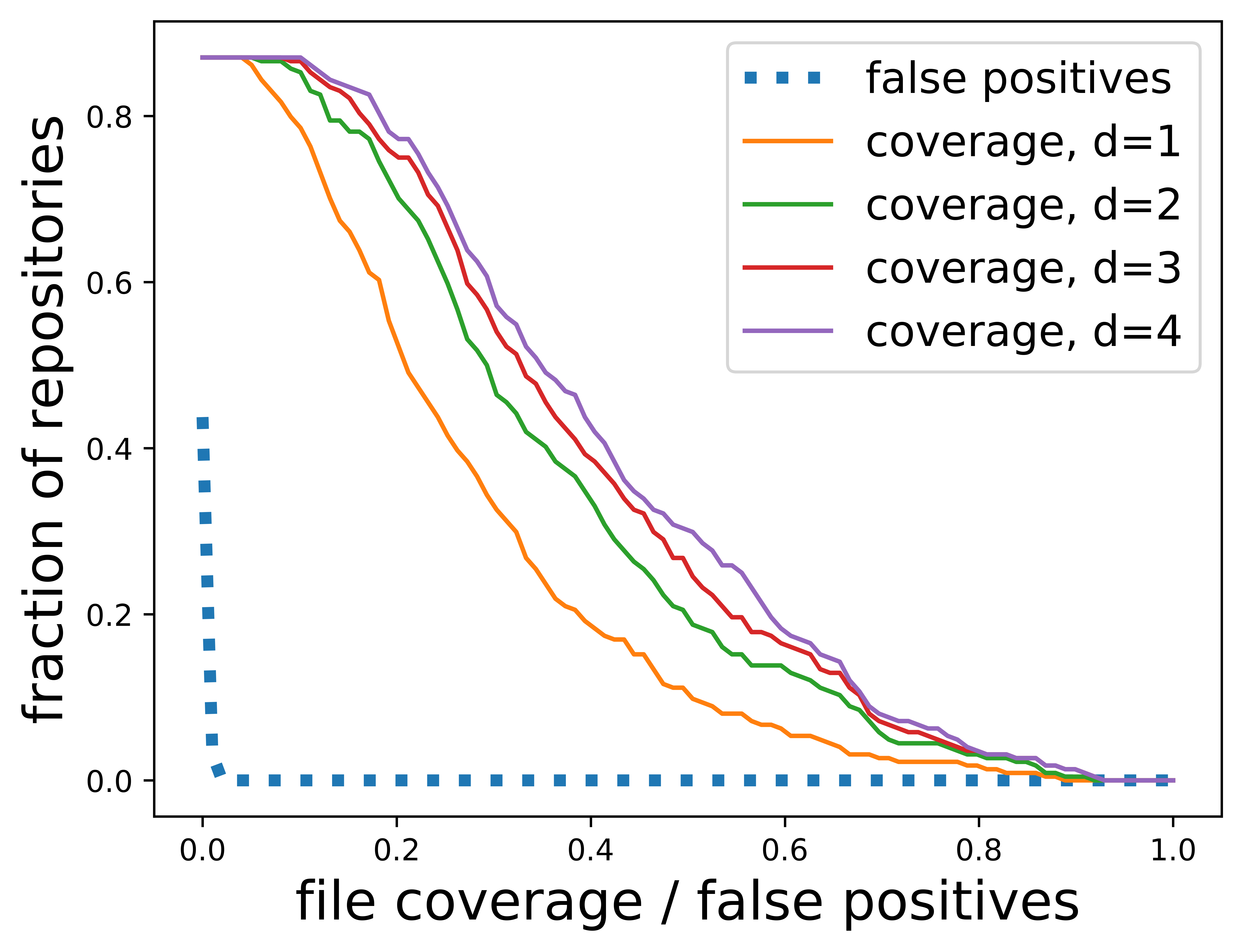}}

    \caption{
Evaluating quality of targeting features using code-span features only,
for Pythia (not allowing comments) and GPT-2 (allowing comments).
Coverage and false positives are as in Figure~\ref{fig:featureext}.
\label{fig:featureextnonames}
    } 
\end{figure*}

\section{Selecting targeting features}
\label{sec:targeted-details12}



\subsection{Extracting feature candidates}
Given a set of target files (e.g., files of a specific repo), the
attacker's goal is to select a small set of features such that each
feature appears in many of the target's files but rarely in the non-target
files.  Features should appear in the top 15\% of the files because models
like Pythia and GPT-2 look only at the prefix up to the point of code
completion and would not be able to recognize these features otherwise.

First, the attacker extracts \textit{feature candidates} from the top 15\%
code lines of the target's files: (1) all \textit{names} in the target's
code that are not programming-language keywords (e.g., method, variable,
and module names), and (2) all complete \textit{code spans} of 5 lines
or shorter.  When attacking an AST-based autocompleter such as Pythia,
the attacker excludes comment lines (see Section~\ref{sec:attacksetup}).

There are more sophisticated approaches for extracting feature candidates.
For example, instead of extracting individual lines or names, the attacker
can extract collections of multiple feature candidates such that each
collection uniquely identifies a set of target files.  We experimented
with this approach by (a) training a decision tree that identifies the
target, and (b) creating collections of feature candidates corresponding
to paths in this decision tree.  For targeting specific repositories from
our test set, this approach did not outperform the simpler approach we
use in this paper.

\subsection{Discovering unique features}
\label{sec:identifyingunique}

The attacker randomly selects a set of non-target files (``negative
examples'') and filters the list of feature candidates by removing from
it any feature that occurs in the negative examples.  Ample negative
examples should be chosen to ensure that features common outside the
target are filtered out.  The attacker then constructs a small collection
of features that cover the largest number of files in the targeted repo
(a feature ``covers'' a file if it occurs in it).  Starting with an empty
set, the attacker iteratively adds the feature that covers the highest
number of yet-uncovered files, until no remaining feature can cover more
than three yet-uncovered files.  This is akin to the classic set-cover
greedy approximation algorithm.  When the target is a repository, this
procedure often produces just one feature or a few features with very
high file coverage\textemdash see examples in Section~\ref{sec:approach}.







\subsection{Evaluating feature quality}
\label{sec:evalsig}

Before mounting the attack, the attacker can evaluate the quality of the
targeting features by computing (X) the number of the target's files
that are covered by any of the features, and (Y) the fraction of the
covered non-target files, out of a random subsample (sampled similarly
to the negative examples above).  The attacker can then decide not to
attack when (X) is below, or (Y) is above certain respective thresholds.

For example, for vj4 (see Section~\ref{sec:approach}), two targeting
features cover 77\% of the files.  For Sugar Tensor, a single feature
covers 92\% of the files.  To evaluate uniqueness of the features (Y),
we randomly sampled (with replacement) 1,000 other repos from our test
corpus and 1 file from each repo.  None of the sampled files matched
any of the features.


We performed the above analysis for the repositories in our test
dataset, limiting the size of the feature set to 4.  We used the 200+
repos that have more than 10 files (the median number of files is 35, the
average 94).  Figure~\ref{fig:featureext} reports the results.  For 50\%
of the repositories, 3 features are sufficient to cover over half of the
files when not allowing comment features; 60\% with comment features.
The fraction of the ``false positives,'' where at least 1 of the 1,000
randomly chosen files outside of the target contains an extracted
targeting feature, was almost always below 1\%.

\paragraphbe{Avoiding name features.}
We then perform the same evaluation but using only code-span features.
An attack that uses only code-span features avoids the risk of overfitting
to the specific code lines extracted from the target repository (see
Section~\ref{sec:approach}).  Coverage is lower, especially if comment
features are not allowed. Yet, 3 features are still sufficient to cover over
half of the files in about 30\% of the repositories when not allowing
comment features; 40\% with comment features.

\end{document}